\documentclass[twocolumn,floatfix,prx,aps,showpacs]{revtex4-1}
\usepackage{graphicx,amsmath,amssymb,color}
\usepackage{nicefrac}
\usepackage{multirow, makecell, ragged2e}
\usepackage[titletoc,title]{appendix}

\newcommand{\be}{\begin{equation}}
\newcommand{\ee}{\end{equation}}

\newcommand{\ba}{\begin{eqnarray}}
\newcommand{\ea}{\end{eqnarray}}

\renewcommand{\vec}[1]{\mbox{\boldmath$#1$}}

\def\beq{\begin{equation}}
\def\eeq{\end{equation}}

\begin{document}

\title{Crystallization in the Fractional Quantum Hall Regime Induced by Landau-level Mixing}
\author{Jianyun Zhao, Yuhe Zhang and J. K. Jain}
\affiliation{Department of Physics, 104 Davey Lab, The Pennsylvania State University, University Park, Pennsylvania 16802}

\date{\today}

\begin{abstract}
{The interplay between strongly correlated liquid and crystal phases for two-dimensional electrons exposed to a high transverse magnetic field is of fundamental interest. Through the non-perturbative fixed phase diffusion Monte Carlo method, we determine the phase diagram of the Wigner crystal in the $\nu-\kappa$ plane, where $\nu$ is the filling factor and $\kappa$ is the strength of Landau level mixing. The phase boundary is seen to exhibit a striking $\nu$ dependence, with the states away from the magic filling factors $\nu=n/(2pn+1)$ being much more susceptible to crystallization due to Landau level mixing than those at $\nu=n/(2pn+1)$.  Our results explain the qualitative difference between the experimental behaviors observed in n-doped and p-doped GaAs quantum wells, and, in particular, the existence of an insulating state for $\nu<1/3$ and also for $1/3 <\nu< 2/5$ in low density p-doped systems.  We predict that in the vicinity of $\nu=1/5$ and $\nu=2/9$, increasing LL mixing causes a transition not into an ordinary electron Wigner crystal but rather into a strongly correlated crystal of composite fermions carrying two vortices. 
}
\end{abstract}

\maketitle

The search for two-dimensional Wigner crystal\cite{Wigner34} in high magnetic fields has led to profound discoveries. The original idea\cite{Lozovik75} was to induce a crystal state of electrons in two dimensions by effectively quenching their kinetic energy with the application of a strong transverse magnetic field, which drives them into the lowest Landau level (LL).  While searching for the Wigner crystal, Tsui, Stormer and Gossard discovered \cite{Tsui82} the $\nu=1/3$ Laughlin liquid\cite{Laughlin83}. As we now know, over a range of filling factors the crystal phase is superseded by the formation of a topological quantum liquid of composite fermions\cite{Jain07,Jain89,Lopez91,Halperin93}, manifesting through fractional quantum Hall (FQH) effect at $\nu=n/(2n\pm 1)$ and $\nu=n/(4n\pm 1)$ and Fermi seas at $\nu=1/2$ and $\nu=1/4$. Theory suggested that the crystal should occur at sufficiently low filling factors\cite{Lam84,Levesque84}, and extensive experimental work has been performed toward determining the phase boundary between the crystal and the liquid\cite{Shayegan07,Fertig07,Jiang90, Goldman90, Paalanen92, Manoharan94a, Engel97, Pan02, Li00, Ye02, Chen04, Csathy05, Sambandamurthy06, Chen06}.  For n-doped GaAs samples, in the limit of zero temperature, an insulating phase is seen for $\nu<1/5$, and also for a narrow range of fillings between $1/5$ and $2/9$. These features have persisted as the sample quality has significantly improved, indicating that the insulator is a pinned crystal rather than an Anderson-type single particle localized state.  Direct evidence for a periodic lattice has been seen through commensurability oscillations \cite{Deng16}.  These  observations are largely understood. Interestingly, theory suggests that at low $\nu$ nature exploits {\em both} the composite fermion (CF) and the crystalline correlations to form a CF crystal\cite{Yi98,Narevich01,Chang05,Archer13} (see Ref.~\cite{Chang05} for a quantitative comparison with the Coulomb ground state) rather than an electron crystal\cite{Maki83,Lam84,Levesque84}.  There is growing experimental support for the CF nature of the crystal\cite{Liu14a,Zhang15c,Jang17}.

A striking puzzle has however persisted since the early 1990s, namely a qualitative difference between the n-doped and p-doped GaAs systems\cite{Santos92,Santos92b,Pan05}. In low-density p-doped GaAs systems, while the FQH states at 1/3 and 2/5 are robust, an insulating phase is observed for filling factors below 1/3, and even between 1/3 and 2/5. In contrast, there is no sign of crystal in this range of $\nu$ in the n-doped samples with the same or even smaller densities. Several early authors\cite{Zhu93, Price93, Platzman93,Ortiz93} attributed this difference to the stronger LL mixing in p-doped GaAs quantum wells due to the larger effective mass of holes, and showed that LL mixing generally favors the crystal phase by studying the competition between the Laughlin liquid and the crystal state at fractions $\nu=1/3$, 1/5 and 1/7 through variational\cite{Zhu93, Price93, Platzman93}, diffusion\cite{Ortiz93}, and path integral Monte Carlo\cite{He05}. More recent experiments in ZnO quantum wells\cite{Maryenko17}, where LL mixing is comparable to that in p-doped GaAs systems, also show insulating phases intermingled with the $\nu=n/(2n+1)$ FQH liquids.

We investigate in this article the competition between liquid and crystal states treating LL mixing non-perturbatively using the fixed phase diffusion Monte Carlo (DMC) method of Ortiz, Ceperley and Martin (OCM) \cite{Ortiz93}.  Two important aspects of our work are: a) we address the issue as a function of continuous filling $\nu$, which is necessary for understanding the observed re-entrant phase transitions; and b) we use accurate crystal and liquid wave functions as the guiding trial wave functions. The FQH state at $\nu=\nu^*/(2\nu^*+1)$ maps into a state of $^2$CFs at filling $\nu^*$, which is in general not an integer. (The symbol $^2$CF refers to composite fermion carrying two quantized vortices.) We assume a model\cite{Archer13} in which the $^2$CFs in the partially filled $\Lambda$ level (i.e. Landau-like level of composite fermions) form a crystal. Although this state has a crystalline order, we refer to it as FQH liquid in the following, because the pinning of this crystal by disorder results in a quantized Hall resistance. (Such a crystal residing on top of a FQH state is called a type-II CF crystal, by analogy to the type-II superconductor which exhibits zero resistance when the Abrikosov flux lattice is pinned.) The insulating state is modeled as a pinned ``type-I" crystal of electrons or composite fermions in which {\em all} particles form a crystal. Extremely accurate lowest LL (LLL) wave functions are available for these states, which we use to fix the phase of the wave function in the DMC method; this is important because the accuracy of the results depends sensitively on the choice of the phase. (We note that while we use the terminology ``electron crystal" or ``electron liquid," our results below apply to both electron and hole systems.)

\begin{figure}
\hspace{-3mm}
\includegraphics[scale=0.25]{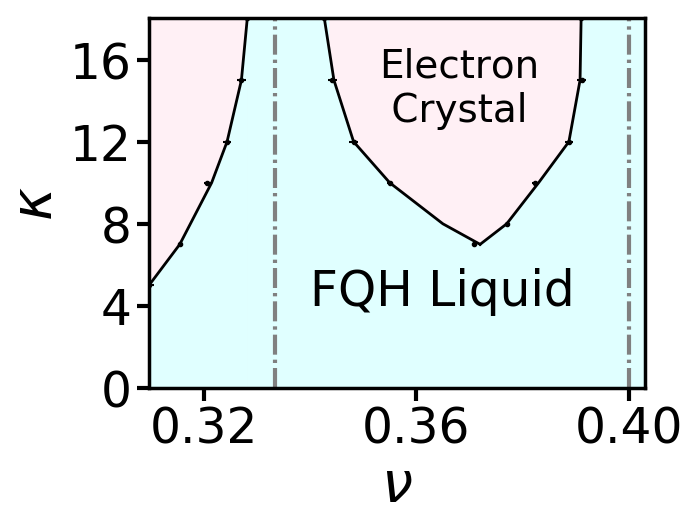}
\hspace{-2.3mm}
\includegraphics[scale=0.25]{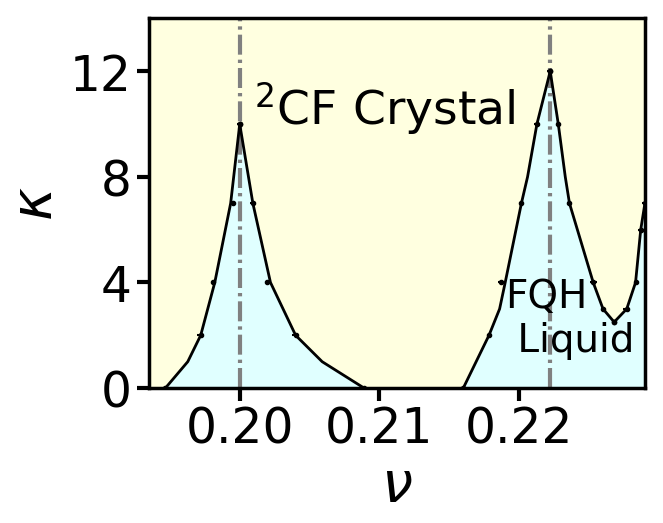}
\caption{
Left panel shows the phase diagram of the electron crystal and the FQH liquid in a filling factor range including $\nu=1/3$ and $\nu=2/5$ as a function of the LL mixing parameter $\kappa$. While the 1/3 and 2/5 FQH states are very robust to LL mixing, for intermediate fillings the crystal appears for $\kappa\gtrsim 7$. The right panel displays the theoretical phase diagram of the $^2$CF crystal and FQH liquid in a range including $\nu=1/5$ and $\nu=2/9$. The electron crystal is not stabilized in this filling factor region because it has substantially higher energy than the $^2$CF crystal\cite{SM-Zhao-2018}.  The uncertainty in the location of the phase boundaries is $\Delta\nu \lesssim$ 0.001 within our model defined in the text.
}
\label{phase}
\end{figure}

Following the usual convention, we quantify the strength of LL mixing through the parameter $\kappa=(e^2/\epsilon l)/(\hbar eB/m_bc)$, which is  the ratio of  the Coulomb energy to the cyclotron energy. (Here, $l=\sqrt{\hbar c/eB}$ is the magnetic length, $m_b$ is the band mass, and $\kappa$ is related to the standard parameter $r_s$ as $\kappa=\sqrt{\nu/2}\;r_s$.) Our principal result is the phase diagrams shown in Fig.~\ref{phase}. The most striking feature they reveal is the strong $\nu$ dependence of the phase boundary separating the FQH and the crystal phases. For example, the FQH effect at $\nu=1/3$ and 2/5 survives up to the largest value of $\kappa$ ($=18$) we have considered, but the electron crystal appears already at $\kappa\gtrsim 7$ for certain $\nu$ in between 1/3 and 2/5, and at even lower values of $\kappa$ for $\nu<1/3$.  Another notable feature is that in the vicinity of $\nu=1/5$ and 2/9, LL mixing induces a transition into the strongly correlated $^2$CF crystal rather than an electron crystal. (If we only considered the {\em electron} crystal, no transition into the crystal state would occur at $\nu=1/5$ and $\nu=2/9$ for up to $\kappa=18$.)  In what follows, we give details of calculations leading to these phase diagrams, and discuss their connection to experiments.

{\em Fixed phase DMC:} The goal is to find the minimum energy $\langle \Psi(\mathcal R)|H|\Psi(\mathcal R) \rangle|$ by varying over the entire Hilbert space of states, where $H$ is the Hamiltonian for interacting two-dimensional electrons in a magnetic field and $\mathcal R$ represents the particle coordinates $\{\vec{r}_j \}$. Because this is not feasible for fermions, we employ an approximate strategy called the fixed phase DMC\cite{Ortiz93} wherein we search for the ground state in a restricted subspace. (The fixed phase DMC is closely related to the fixed node DMC.\cite{Melton17}) Following OCM, we substitute $\Psi(\mathcal R)=\Phi(\mathcal R)e^{i\varphi(\mathcal R)}$ where $\Phi(\mathcal R)=|\Psi(\mathcal R)|$ is real and non-negative. The above energy is then given by $\langle \Phi(\mathcal R)|H_R|\Phi(\mathcal R) \rangle|$ with $H_R= \sum_{j=1}^N \left[ \vec{p}_j^2 +[\hbar\vec{\nabla}_j\varphi(\mathcal R)+(e/c)\vec{A}(\vec{r}_j)]^2\right]/2m+V_{\rm Coulomb}(\mathcal R)$. Now, keeping the phase $\varphi(\mathcal R)$ fixed and varying $\Phi(\mathcal R)$ gives us the lowest energy within the subspace of wave functions defined by the phase sector $\varphi(\mathcal R)$. This minimization is most conveniently accomplished by the DMC method\cite{Reynolds82,Foulkes01}. In this approach, one views the imaginary time Schr\"odinger equation, $-\hbar{\partial\over\partial t}\Phi(\mathcal R,t)=\left[ H_R(\mathcal R)-E_T)\right]\Phi(\mathcal R,t)$, as a diffusion equation, where $\Phi(\mathcal R,t)$ is interpreted as the probability distribution of the diffusing ``walkers" and $E_T$ is an energy offset. Evolving this equation in imaginary time projects out the lowest energy state, which is the ground state provided that the initial trial wave function has a non-zero overlap with the ground state. DMC is a method for implementing this scheme through importance sampling, where ``walkers" in the $2N$ dimensional configuration space proliferate (die) in regions of low (high) potential energy according to certain standard rules, and converge into the probability distribution of the ground state in the limit $t\rightarrow\infty$. The fixed phase DMC produces the lowest energy in the chosen phase sector, and hence a variational upper bound for the exact ground state energy.

We  perform our calculations in the spherical geometry\cite{Haldane83} in which electrons are confined on the surface of a sphere, with a flux $2Q\phi_0$ passing radially through it, where $2Q$ is an integer and $\phi_0=hc/e$ is the flux quantum.  We  use $l$ as the unit of length and  $e^2\over\epsilon l$ as the unit of energy.  The particle position is identified through the ``spinor" coordinates $u={\rm cos}(\theta/2)e^{i\phi/2}$ and $v={\rm sin}(\theta/2)e^{-i\phi/2}$. Melik-Alaverdian, Bonesteel and Ortiz \cite{Melik-Alaverdian97} have formulated the fixed phase DMC in the spherical geometry through a stereographic projection, and we will follow their method.

\begin{figure}
\hspace{-6mm}
\includegraphics[scale=0.19]{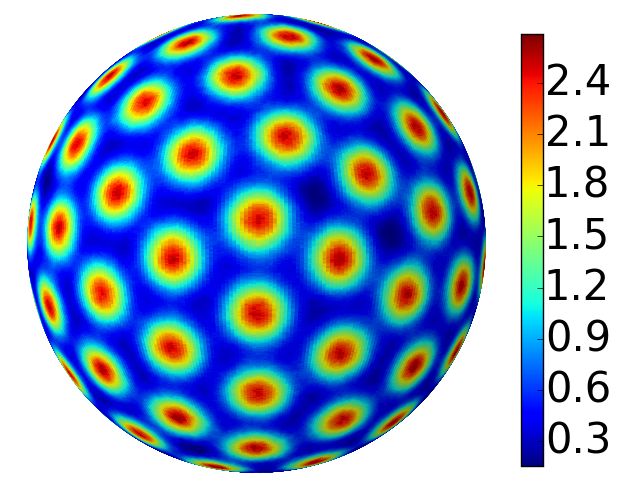}
\hspace{-2.6mm}
\includegraphics[scale=0.19]{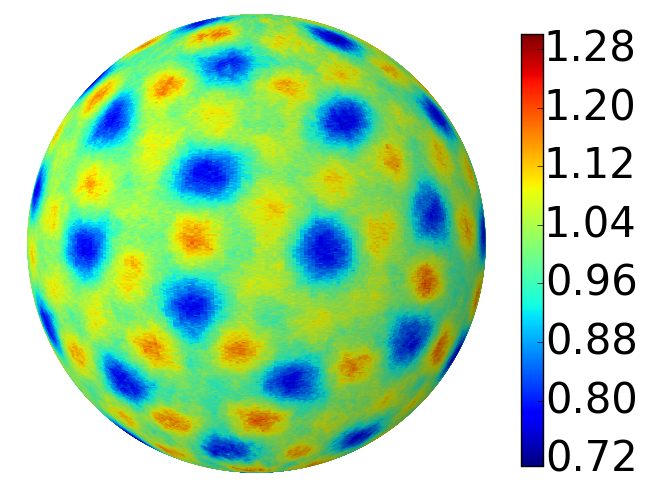}
\hspace{-2.6mm}
\includegraphics[scale=0.19]{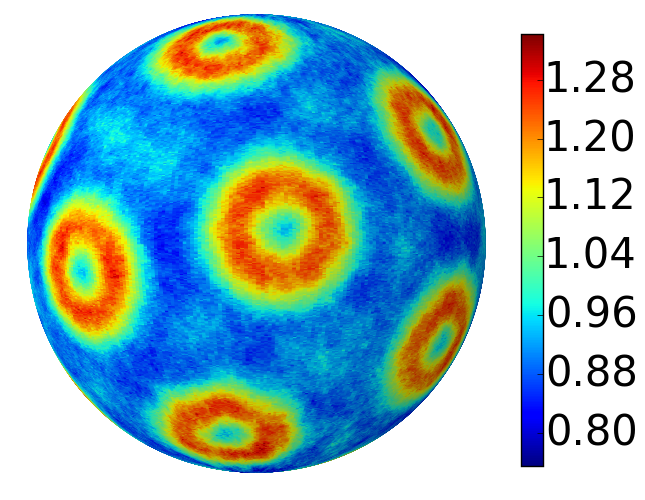}
\caption{
Density profiles of various crystals for $N=96$ particles at severals fillings. Left shows a type-I electron crystal for $\nu=0.394$ ($2Q=240$), and the middle shows a type-II CF crystal for the same parameters, and right panel shows a type-II CF crystal for $\nu=0.351$ ($2Q=270$). The density is given in units of the average density. All results are for $\kappa=0$.
}
\label{structure}
\end{figure}

{\em Trial wave functions:}
The accuracy of the energies obtained from fixed phase DMC is critically dependent on the choice of the phase $\varphi(\mathcal R)$.  G\"u\c{c}l\"u and Umrigar~\cite{Guclu05} found in studies of certain small systems that the phase of the wave function is not significantly altered by LL mixing. Following their lead, we shall use the accurate LLL wave functions as the trial wave functions to fix the phase $\varphi(\mathcal R)$ in our calculations.  In cases where we have been able to compare (e.g. the $^2$CF crystal vs. the Hartree-Fock crystal in the vicinity of $\nu=1/5$; the CF Fermi sea vs. the Pfaffian wave function at $\nu=1/2$; the LLL projected vs. the unprojected FQH wave functions) we have found that fixing the phase with the more accurate LLL wave function produces lower energy for up to the largest values of $\kappa$ we considered.  Nonetheless, our results are subject to our assumption regarding the phase, the validity of which can ultimately be justified only by a detailed comparison of our results with experiments. We note that this method of phase fixing has yielded a decent quantitative account of spin phase transitions\cite{Zhang16,Zhang17}.

In the spherical geometry, a localized wave packet centered at $(U,V)$ is given by $(U^*u+V^*v)^{2Q}$ where $(u,v)$ are particle coordinates. (This is the delta function projected in the LLL.) The wave function for the type-I electron crystal \cite{Maki83} is given by $\det(U_l^*u_i+V_l^*v_i)^{2Q}$, and for the type-I $^{2p}$CF crystal by:
\beq
\Psi^\textrm{CFC}_{2Q} =  
\prod_{j<k}(u_jv_k-v_ju_k)^{2p} \det(U_l^*u_i+V_l^*v_i)^{2Q^*}
\label{2pCFC}
\eeq
where $2Q^*=2Q-2p(N-1)$ and $(U_l, V_l)$ are the spherical coordinates for the crystal lattice sites. Because it is not possible to fit a hexagonal lattice perfectly on the surface of a sphere, we choose our crystal sites that minimize the Coulomb energy of point charges on a sphere. This is the famous Thomson problem\cite{Thomson04}, and the positions have been evaluated numerically and available in the literature\cite{Wales06,Wales09,Thomson}. As expected, the Thomson lattice has a triangular structure locally but contains some defects. We  consider the thermodynamic limit to eliminate the contribution from defects. We also note that only those values of $2p$ are allowed that produce a positive value for $2Q^*$. In particular, for $\nu\geq 1/3$, we only have the $2p=0$ electron crystal available (at $1/3$, the $^2$CF crystal wave function in Eq.~\ref{2pCFC} becomes identical to the Laughlin liquid wave function) and for $\nu<1/3$ we can form crystals also with $2p=2$.

For the FQH state of electrons at flux $2Q$ we construct $\Psi_{2Q}=\mathcal P_{\rm{LLL}}\Phi_{2Q^*}\prod_{j<k}(u_jv_k-v_ju_k)^{2p}$, where $\Phi_{2Q^*}$ is the wave function of electrons at effective flux $2Q^*=2Q-2p(N-1)$ and $\mathcal P_{\rm{LLL}}$ is the LLL projection operator, which will be evaluated using standard methods\cite{Jain97,Jain97b}. When $2Q^*$ corresponds to an integer filling $\nu^*=n$, we obtain wave function for electrons at $\nu=n/(2pn+1)$. When $\nu^*>1$, we will assume that the composite fermions in the topmost partially filled level form a Thomson crystal, and for $\nu^*<1$, we will assume a crystal of CF holes in the lowest $\Lambda$ level. This assumption is expected to be accurate when the density of CF particles or holes is small, and a good first approximation in the entire $\nu$ range we have considered. The density profiles for certain type-I and type-II crystals on the surface of a sphere are shown in Fig.~\ref{structure}.

\begin{figure}
\hspace{-5mm}
\includegraphics[scale=0.33]{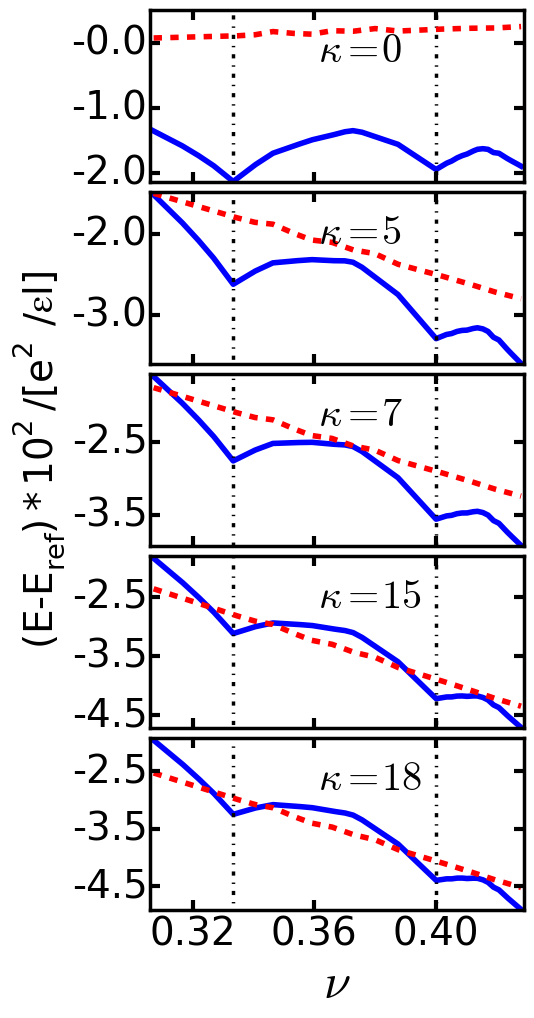}
\hspace{-2.5mm}
\includegraphics[scale=0.33]{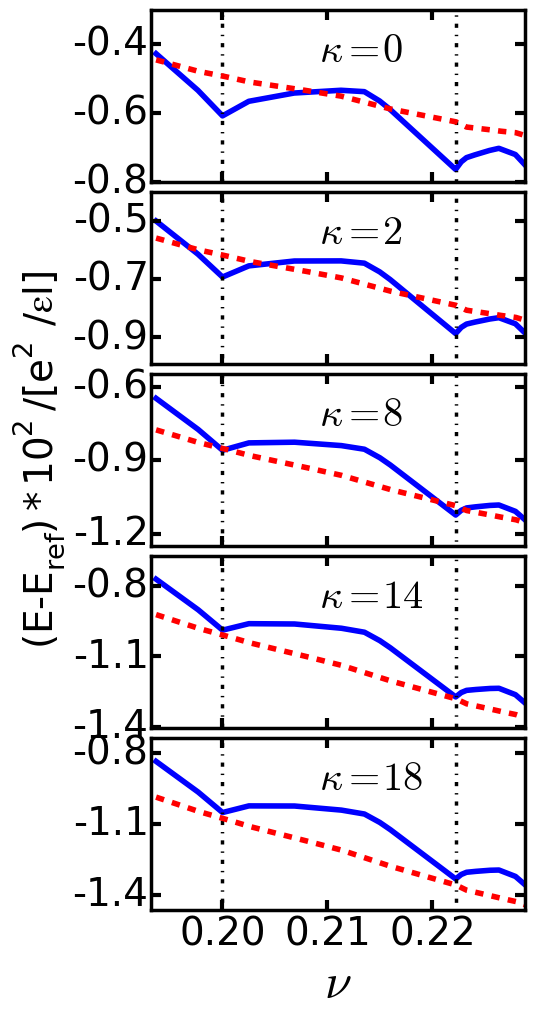}
\caption{
Energies of the FQH state (blue line) and the crystal (red dashed line) as a function of the filling factor in the vicinity of $\nu=1/3$ (left columns) and $\nu=1/5$ (right columns). The quoted energies are measured relative to $E_{\rm ref}=-0.782133\nu^{1/2}+0.2823\nu^{3/2}+0.18\nu^{5/2}-1.41e^{-2.07/\nu }$, which is the energy of a Wigner crystal in the Hartree-Fock approximation\cite{Lam84}. 
}
\label{energy_comp}
\end{figure}

{\em Results:} In the spherical geometry the relation between $\nu$, $N$ and $2Q$ has the form $2Q=\nu^{-1}N-S_\nu$, where $S_\nu$ is called the ``shift." We define the filling factor as $\nu={N+2\over2Q+3(2p+1)}$ which gives the correct shifts \cite{Jain07} of $S_\nu=2p+1$ and $S_\nu=2p+2$ at $\nu=1/(2p+1)$ and $\nu=2/(4p+1)$ and is sufficient for our considerations. All energies quoted below are energies per particle, and are corrected for the fact that the density in the spherical geometry has an $N$ dependent deviation from the thermodynamic density; this corresponds to multiplication by $\sqrt{2Q\nu\over N}$. We find that the density correction makes the energies $N$ independent to a large extent.

It is crucial to obtain the thermodynamic value for the ground state energy of the electron or the $^{2p}$CF crystal. This is complicated by the fact that the fixed phase DMC energy for $N\lesssim 35$, the only systems accessible to our fixed phase DMC calculation, shows substantial finite size fluctuations due to the inevitable presence of defects, thereby precluding a reliable extrapolation to the thermodynamic limit. (See the Supplemental Materials\cite{SM-Zhao-2018} for details.) 
Fortunately, we find\cite{SM-Zhao-2018} that the energy {\em difference} $\Delta E_{\kappa}^N(\nu)\equiv E_{\kappa}^N(\nu)-E_{\kappa=0}^N(\nu)$ is very well behaved and nearly constant as a function of $N$, leading to an accurate thermodynamic value $\Delta E_{\kappa}(\nu)$ using systems with up to $N\lesssim 35$. Furthermore, it is possible to obtain the thermodynamic limits of $E_{\kappa=0}(\nu)$ very precisely for $\kappa=0$, because here we only need to perform variational Monte Carlo and can access much larger $N$. The quantity $E_{\kappa}(\nu)=E_{\kappa=0}(\nu)+\Delta E_{\kappa}(\nu)$ thus produces an accurate value for the CF crystal energy for non-zero $\kappa$. We use the same method to obtain the energy of the FQH liquid phase.

Fig.~\ref{energy_comp} shows the energies of the liquid and crystal states for a system with 96 particles, which is large enough that the results reflect the thermodynamic limit. For this purpose, we first obtain the energies of the liquid and crystal states by variational Monte Carlo at $\kappa=0$, and then add $\Delta E_{\kappa}(\nu)$ to it to obtain the values shown in the figure. To obtain the energy reduction $\Delta E_{\kappa}(\nu)$ due to LL mixing, we assume that $\Delta E_{\kappa}(\nu)$ is a smooth function of $\kappa$ in the narrow filling factor ranges considered and therefore it is sufficient to determine $\Delta E_{\kappa}(\nu)$ only for $\nu_1=1/(2p+1)$ and $\nu_2=2/(4p+1)$, and then use $\Delta E_{\kappa}(\nu)={\Delta E_{\kappa}(\nu_2)-\Delta E_{\kappa}(\nu_1)\over\nu_2-\nu_1}(\nu-\nu_1) + \Delta E_{\kappa}(\nu_1)$ at arbitrary $\nu$ in the neighborhood. The phase diagrams in Fig.~\ref{phase} are obtained from the crossing points, which are determined with an uncertainty of $\Delta\nu \lesssim$ 0.001 within our model.

\begin{figure}
\includegraphics[scale=0.23]{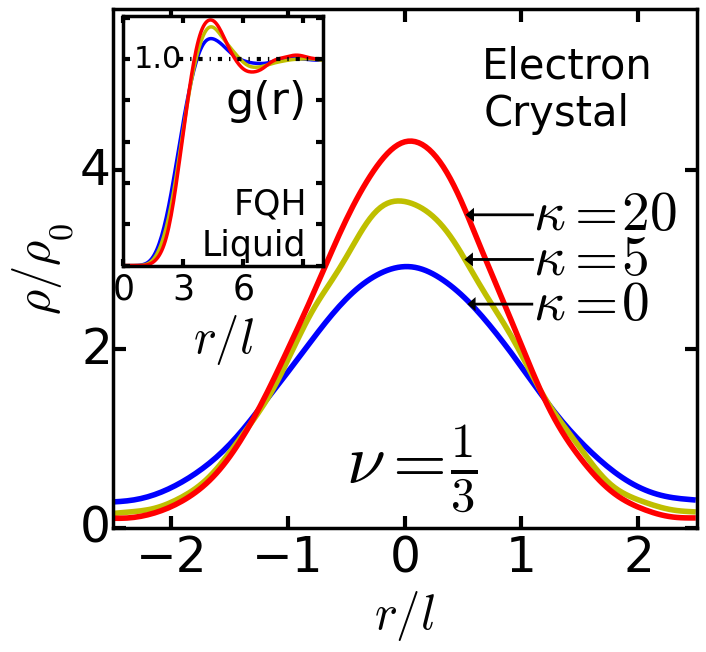}
\includegraphics[scale=0.23]{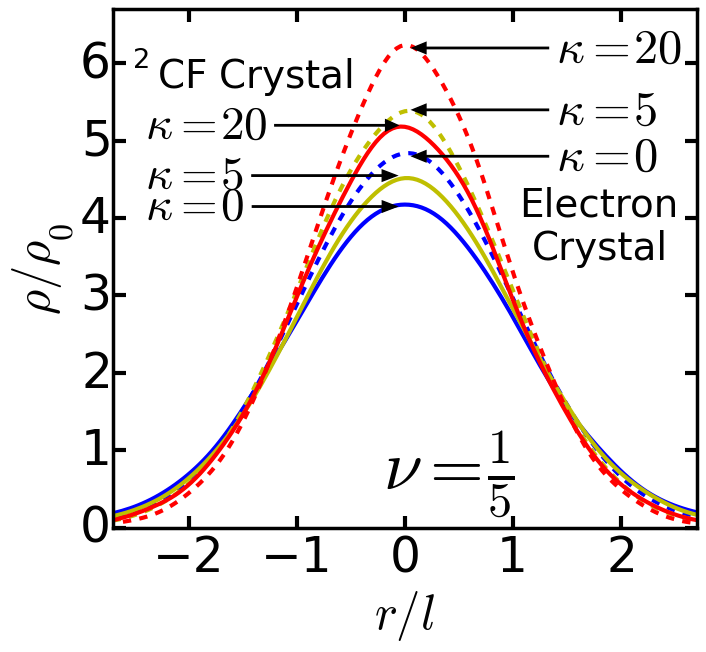}
\caption{
The left panel depicts the density of the localized wave packet at a crystal site as a function of $\kappa$ for $\nu=1/3$. In the right panel, the density profiles at a crystal site are shown $\nu=1/5$ for both the electron crystal (dashed lines) and $^2$CF crystal (solid lines). All densities are quoted in units of $\rho_0$, the density of one full LL. The $\kappa$ dependence for the pair correlation function at $\nu=1/3$ is shown as an inset in the left panel. The blue, yellow and red curves correspond to $\kappa=0$, $\kappa=5$, and $\kappa=20$, respectively. The results are for a system of $N=24$ particles.
}
\label{peak}
\end{figure}

One may ask why the crystal phase is favored by LL mixing.  LL mixing allows the electron wave packet at each site to become more localized, which enhances the kinetic energy but reduces the interaction energy. In Fig.~\ref{peak} we show the line shape of the localized wave packet as a function of $\kappa$ for both the electron and the $^2$CF crystals.  Fig.~\ref{peak} also shows how the pair correlation function of the liquid changes as a function of $\kappa$. [The pair correlation function is evaluated by the so-called mixed-estimator (see, for example, Ref.~\cite{Foulkes01}): $g(\vec r)=\rho_0^{-2}\left\langle\Phi\middle|\hat\rho(\vec 0)\hat\rho(\vec r)\middle|\Phi_T\right\rangle/\left\langle\Phi\middle|\Phi_T\right\rangle$, where $\Phi_T$ is the initial trial state, $\Phi$ is the ``final" ground state, and $\rho_0$ is the average density.]  While the energies of both the liquid and crystal states are reduced, only a detailed and quantitatively reliable calculation can tell if and where a transition into a crystal takes place.

{\em Comparison with Experiment:}  In n-type GaAs quantum wells, with $\epsilon=12.5$ and $m_b=0.067m_e$ we have $\kappa\approx 2.6/\sqrt{B[T]}\approx1.28\sqrt{\nu/(\rho/10^{11}\rm{cm}^{-2})}$. For $\nu=1/3$ ($\nu=1/5$) we have $\kappa=0.74$ ($\kappa=0.57$) for $\rho=1.0\times 10^{11}$ cm$^{-2}$ and $\kappa=2.3$ ($\kappa=1.8$) for $\rho=1.0\times 10^{10}$ cm$^{-2}$. For these values, both the 1/3 and 1/5 states are deep in the FQH regime.  The same is true of 2/5 and 2/9. This is consistent with the observation that all best quality n-doped samples show these FQH states. Furthermore, an insulator is seen in between 1/5 and 2/9 as well as below 1/5 in all high quality samples. 

For p-doped samples, the larger value of $\kappa$ makes the situation more interesting.  
The $\kappa$ for holes in GaAs is $\approx 5.6$ times that for electrons for the same $B$ \cite{Sodemann13}.  For $\nu=1/3$ the p-doped samples have $\kappa\approx 4$ for hole density $\rho=1.0\times 10^{11}$ cm$^{-2}$ and $\kappa\approx 13$ for $\rho=1.0\times 10^{10}$ cm$^{-2}$. This implies that for low density p-doped samples, an insulating crystal state can occur in between 1/3 and 2/5 and also below $\nu=1/3$, although the 1/3 and 2/5 state should remain FQH liquids. At $\nu\approx 0.37$ (which lies between 1/3 and 2/5), $\kappa=7$ corresponds to $\rho \approx 4 \times 10^{10}$ cm$^{-2}$. In Santos {\em et al.}\cite{Santos92b} a transition is seen at $\rho\approx 7\times 10^{10}$ cm$^{-2}$. Much lower values of $\kappa$ are required to produce a crystal for $\nu<1/3$. Theoretical predictions are thus in qualitative and good semi-quantitative agreement with experimental observations.  Similar considerations apply to ZnO quantum wells \cite{Maryenko17} for which $\kappa$ is $\sim 6.4$ times larger than that for n-doped GaAs systems\cite{Sodemann13}.

One may ask to what extent the tiny energy differences at the rather large values of $\kappa$ are affected by the choice of our phase. While this issue deserves further examination, the qualitative and semi-quantitative agreement with experiment lends some degree of a posteriori justification for the phase choice.  

We have carried out similar calculations assuming a quantum well of width $w$ with a transverse wave function $\xi(\eta)=\sqrt{2 /w}\sin{\pi \eta/ w}$, where the transverse coordinate $0<\eta<w$.  
The effective 2D interaction is then given by $V^{\rm eff}(r)=\int_0^w d\eta_1\int_0^w d\eta_2 {|\xi(\eta_1)|^2 |\xi(\eta_2)|^2\over [r^2+(\eta_1-\eta_2)^2]^{1/2}}$ where $r=\sqrt{(x_1-x_2)^2+(y_1-y_2)^2}$.  
Our calculations predict\cite{SM-Zhao-2018} that the transition into the crystal state is pushed to approximately 20\% higher $\kappa$ values for $w=2l$. 
For large widths (e.g. $w=4l$), the wave function develops double hump structure\cite{Manoharan96}, which we have not considered in this work for simplicity.

We are grateful to Ajit Balram and Mansour Shayegan for very useful discussions, and acknowledge financial support from the DOE Grant No. DE-SC0005042.

%\bibliography{../../Latex-Revtex-etc./biblio_fqhe.bib}

\begin{thebibliography}{59}
\expandafter\ifx\csname natexlab\endcsname\relax\def\natexlab#1{#1}\fi
\expandafter\ifx\csname bibnamefont\endcsname\relax
  \def\bibnamefont#1{#1}\fi
\expandafter\ifx\csname bibfnamefont\endcsname\relax
  \def\bibfnamefont#1{#1}\fi
\expandafter\ifx\csname citenamefont\endcsname\relax
  \def\citenamefont#1{#1}\fi
\expandafter\ifx\csname url\endcsname\relax
  \def\url#1{\texttt{#1}}\fi
\expandafter\ifx\csname urlprefix\endcsname\relax\def\urlprefix{URL }\fi
\providecommand{\bibinfo}[2]{#2}
\providecommand{\eprint}[2][]{\url{#2}}

\bibitem[{\citenamefont{Wigner}(1934)}]{Wigner34}
\bibinfo{author}{\bibfnamefont{E.}~\bibnamefont{Wigner}},
  \bibinfo{journal}{Phys. Rev.} \textbf{\bibinfo{volume}{46}},
  \bibinfo{pages}{1002} (\bibinfo{year}{1934}).

\bibitem[{\citenamefont{Lozovik and Yudson}(1975)}]{Lozovik75}
\bibinfo{author}{\bibfnamefont{Y.~E.} \bibnamefont{Lozovik}} \bibnamefont{and}
  \bibinfo{author}{\bibfnamefont{V.~I.} \bibnamefont{Yudson}},
  \bibinfo{journal}{JETP Lett.} \textbf{\bibinfo{volume}{22}},
  \bibinfo{pages}{11} (\bibinfo{year}{1975}).

\bibitem[{\citenamefont{Tsui et~al.}(1982)\citenamefont{Tsui, Stormer, and
  Gossard}}]{Tsui82}
\bibinfo{author}{\bibfnamefont{D.~C.} \bibnamefont{Tsui}},
  \bibinfo{author}{\bibfnamefont{H.~L.} \bibnamefont{Stormer}},
  \bibnamefont{and} \bibinfo{author}{\bibfnamefont{A.~C.}
  \bibnamefont{Gossard}}, \bibinfo{journal}{Phys. Rev. Lett.}
  \textbf{\bibinfo{volume}{48}}, \bibinfo{pages}{1559} (\bibinfo{year}{1982}),
  \urlprefix\url{http://link.aps.org/doi/10.1103/PhysRevLett.48.1559}.

\bibitem[{\citenamefont{Laughlin}(1983)}]{Laughlin83}
\bibinfo{author}{\bibfnamefont{R.~B.} \bibnamefont{Laughlin}},
  \bibinfo{journal}{Phys. Rev. Lett.} \textbf{\bibinfo{volume}{50}},
  \bibinfo{pages}{1395} (\bibinfo{year}{1983}),
  \urlprefix\url{http://link.aps.org/doi/10.1103/PhysRevLett.50.1395}.

\bibitem[{\citenamefont{Jain}(2007)}]{Jain07}
\bibinfo{author}{\bibfnamefont{J.~K.} \bibnamefont{Jain}},
  \emph{\bibinfo{title}{Composite Fermions}} (\bibinfo{publisher}{Cambridge
  University Press, New York, US}, \bibinfo{year}{2007}).

\bibitem[{\citenamefont{Jain}(1989)}]{Jain89}
\bibinfo{author}{\bibfnamefont{J.~K.} \bibnamefont{Jain}},
  \bibinfo{journal}{Phys. Rev. Lett.} \textbf{\bibinfo{volume}{63}},
  \bibinfo{pages}{199} (\bibinfo{year}{1989}),
  \urlprefix\url{http://link.aps.org/doi/10.1103/PhysRevLett.63.199}.

\bibitem[{\citenamefont{Lopez and Fradkin}(1991)}]{Lopez91}
\bibinfo{author}{\bibfnamefont{A.}~\bibnamefont{Lopez}} \bibnamefont{and}
  \bibinfo{author}{\bibfnamefont{E.}~\bibnamefont{Fradkin}},
  \bibinfo{journal}{Phys. Rev. B} \textbf{\bibinfo{volume}{44}},
  \bibinfo{pages}{5246} (\bibinfo{year}{1991}),
  \urlprefix\url{http://link.aps.org/doi/10.1103/PhysRevB.44.5246}.

\bibitem[{\citenamefont{Halperin et~al.}(1993)\citenamefont{Halperin, Lee, and
  Read}}]{Halperin93}
\bibinfo{author}{\bibfnamefont{B.~I.} \bibnamefont{Halperin}},
  \bibinfo{author}{\bibfnamefont{P.~A.} \bibnamefont{Lee}}, \bibnamefont{and}
  \bibinfo{author}{\bibfnamefont{N.}~\bibnamefont{Read}},
  \bibinfo{journal}{Phys. Rev. B} \textbf{\bibinfo{volume}{47}},
  \bibinfo{pages}{7312} (\bibinfo{year}{1993}),
  \urlprefix\url{http://link.aps.org/doi/10.1103/PhysRevB.47.7312}.

\bibitem[{\citenamefont{Lam and Girvin}(1984)}]{Lam84}
\bibinfo{author}{\bibfnamefont{P.~K.} \bibnamefont{Lam}} \bibnamefont{and}
  \bibinfo{author}{\bibfnamefont{S.~M.} \bibnamefont{Girvin}},
  \bibinfo{journal}{Phys. Rev. B} \textbf{\bibinfo{volume}{30}},
  \bibinfo{pages}{473} (\bibinfo{year}{1984}).

\bibitem[{\citenamefont{Levesque et~al.}(1984)\citenamefont{Levesque, Weis, and
  MacDonald}}]{Levesque84}
\bibinfo{author}{\bibfnamefont{D.}~\bibnamefont{Levesque}},
  \bibinfo{author}{\bibfnamefont{J.~J.} \bibnamefont{Weis}}, \bibnamefont{and}
  \bibinfo{author}{\bibfnamefont{A.~H.} \bibnamefont{MacDonald}},
  \bibinfo{journal}{Phys. Rev. B} \textbf{\bibinfo{volume}{30}},
  \bibinfo{pages}{1056} (\bibinfo{year}{1984}).

\bibitem[{\citenamefont{Shayegan}(2007)}]{Shayegan07}
\bibinfo{author}{\bibfnamefont{M.}~\bibnamefont{Shayegan}}, in
  \emph{\bibinfo{booktitle}{Perspectives in Quantum Hall Effects}}
  (\bibinfo{publisher}{Wiley-VCH Verlag GmbH}, \bibinfo{year}{2007}), p.
  \bibinfo{pages}{343Ð384}, ISBN \bibinfo{isbn}{9783527617258},
  \urlprefix\url{http://dx.doi.org/10.1002/9783527617258.ch10}.

\bibitem[{\citenamefont{Fertig}(2007)}]{Fertig07}
\bibinfo{author}{\bibfnamefont{H.~A.} \bibnamefont{Fertig}}, in
  \emph{\bibinfo{booktitle}{Perspectives in Quantum Hall Effects}}
  (\bibinfo{publisher}{Wiley-VCH Verlag GmbH}, \bibinfo{year}{2007}), p.
  \bibinfo{pages}{71Ð108}, ISBN \bibinfo{isbn}{9783527617258},
  \urlprefix\url{http://dx.doi.org/10.1002/9783527617258.ch10}.

\bibitem[{\citenamefont{Jiang et~al.}(1990)\citenamefont{Jiang, Willett,
  Stormer, Tsui, Pfeiffer, and West}}]{Jiang90}
\bibinfo{author}{\bibfnamefont{H.~W.} \bibnamefont{Jiang}},
  \bibinfo{author}{\bibfnamefont{R.~L.} \bibnamefont{Willett}},
  \bibinfo{author}{\bibfnamefont{H.~L.} \bibnamefont{Stormer}},
  \bibinfo{author}{\bibfnamefont{D.~C.} \bibnamefont{Tsui}},
  \bibinfo{author}{\bibfnamefont{L.~N.} \bibnamefont{Pfeiffer}},
  \bibnamefont{and} \bibinfo{author}{\bibfnamefont{K.~W.} \bibnamefont{West}},
  \bibinfo{journal}{Phys. Rev. Lett.} \textbf{\bibinfo{volume}{65}},
  \bibinfo{pages}{633} (\bibinfo{year}{1990}).

\bibitem[{\citenamefont{Goldman et~al.}(1990)\citenamefont{Goldman, Santos,
  Shayegan, and Cunningham}}]{Goldman90}
\bibinfo{author}{\bibfnamefont{V.~J.} \bibnamefont{Goldman}},
  \bibinfo{author}{\bibfnamefont{M.}~\bibnamefont{Santos}},
  \bibinfo{author}{\bibfnamefont{M.}~\bibnamefont{Shayegan}}, \bibnamefont{and}
  \bibinfo{author}{\bibfnamefont{J.~E.} \bibnamefont{Cunningham}},
  \bibinfo{journal}{Phys. Rev. Lett.} \textbf{\bibinfo{volume}{65}},
  \bibinfo{pages}{2189} (\bibinfo{year}{1990}).

\bibitem[{\citenamefont{Paalanen et~al.}(1992)\citenamefont{Paalanen, Willett,
  Ruel, Littlewood, West, and Pfeiffer}}]{Paalanen92}
\bibinfo{author}{\bibfnamefont{M.~A.} \bibnamefont{Paalanen}},
  \bibinfo{author}{\bibfnamefont{R.~L.} \bibnamefont{Willett}},
  \bibinfo{author}{\bibfnamefont{R.~R.} \bibnamefont{Ruel}},
  \bibinfo{author}{\bibfnamefont{P.~B.} \bibnamefont{Littlewood}},
  \bibinfo{author}{\bibfnamefont{K.~W.} \bibnamefont{West}}, \bibnamefont{and}
  \bibinfo{author}{\bibfnamefont{L.~N.} \bibnamefont{Pfeiffer}},
  \bibinfo{journal}{Phys. Rev. B} \textbf{\bibinfo{volume}{45}},
  \bibinfo{pages}{13784} (\bibinfo{year}{1992}),
  \urlprefix\url{http://link.aps.org/doi/10.1103/PhysRevB.45.13784}.

\bibitem[{\citenamefont{Manoharan and Shayegan}(1994)}]{Manoharan94a}
\bibinfo{author}{\bibfnamefont{H.~C.} \bibnamefont{Manoharan}}
  \bibnamefont{and} \bibinfo{author}{\bibfnamefont{M.}~\bibnamefont{Shayegan}},
  \bibinfo{journal}{Phys. Rev. B} \textbf{\bibinfo{volume}{50}},
  \bibinfo{pages}{17662} (\bibinfo{year}{1994}).

\bibitem[{\citenamefont{Engel et~al.}(1997)\citenamefont{Engel, Li, Shahar,
  Tsui, and Shayegan}}]{Engel97}
\bibinfo{author}{\bibfnamefont{L.}~\bibnamefont{Engel}},
  \bibinfo{author}{\bibfnamefont{C.-C.} \bibnamefont{Li}},
  \bibinfo{author}{\bibfnamefont{D.}~\bibnamefont{Shahar}},
  \bibinfo{author}{\bibfnamefont{D.}~\bibnamefont{Tsui}}, \bibnamefont{and}
  \bibinfo{author}{\bibfnamefont{M.}~\bibnamefont{Shayegan}},
  \bibinfo{journal}{Physica E} \textbf{\bibinfo{volume}{1}},
  \bibinfo{pages}{111 } (\bibinfo{year}{1997}).

\bibitem[{\citenamefont{Pan et~al.}(2002)\citenamefont{Pan, Stormer, Tsui,
  Pfeiffer, Baldwin, and West}}]{Pan02}
\bibinfo{author}{\bibfnamefont{W.}~\bibnamefont{Pan}},
  \bibinfo{author}{\bibfnamefont{H.~L.} \bibnamefont{Stormer}},
  \bibinfo{author}{\bibfnamefont{D.~C.} \bibnamefont{Tsui}},
  \bibinfo{author}{\bibfnamefont{L.~N.} \bibnamefont{Pfeiffer}},
  \bibinfo{author}{\bibfnamefont{K.~W.} \bibnamefont{Baldwin}},
  \bibnamefont{and} \bibinfo{author}{\bibfnamefont{K.~W.} \bibnamefont{West}},
  \bibinfo{journal}{Phys. Rev. Lett.} \textbf{\bibinfo{volume}{88}},
  \bibinfo{pages}{176802} (\bibinfo{year}{2002}).

\bibitem[{\citenamefont{Li et~al.}(2000)\citenamefont{Li, Yoon, Engel, Shahar,
  Tsui, and Shayegan}}]{Li00}
\bibinfo{author}{\bibfnamefont{C.-C.} \bibnamefont{Li}},
  \bibinfo{author}{\bibfnamefont{J.}~\bibnamefont{Yoon}},
  \bibinfo{author}{\bibfnamefont{L.~W.} \bibnamefont{Engel}},
  \bibinfo{author}{\bibfnamefont{D.}~\bibnamefont{Shahar}},
  \bibinfo{author}{\bibfnamefont{D.~C.} \bibnamefont{Tsui}}, \bibnamefont{and}
  \bibinfo{author}{\bibfnamefont{M.}~\bibnamefont{Shayegan}},
  \bibinfo{journal}{Phys. Rev. B} \textbf{\bibinfo{volume}{61}},
  \bibinfo{pages}{10905} (\bibinfo{year}{2000}).

\bibitem[{\citenamefont{Ye et~al.}(2002)\citenamefont{Ye, Engel, Tsui, Lewis,
  Pfeiffer, and West}}]{Ye02}
\bibinfo{author}{\bibfnamefont{P.~D.} \bibnamefont{Ye}},
  \bibinfo{author}{\bibfnamefont{L.~W.} \bibnamefont{Engel}},
  \bibinfo{author}{\bibfnamefont{D.~C.} \bibnamefont{Tsui}},
  \bibinfo{author}{\bibfnamefont{R.~M.} \bibnamefont{Lewis}},
  \bibinfo{author}{\bibfnamefont{L.~N.} \bibnamefont{Pfeiffer}},
  \bibnamefont{and} \bibinfo{author}{\bibfnamefont{K.}~\bibnamefont{West}},
  \bibinfo{journal}{Phys. Rev. Lett.} \textbf{\bibinfo{volume}{89}},
  \bibinfo{pages}{176802} (\bibinfo{year}{2002}).

\bibitem[{\citenamefont{Chen et~al.}(2004)\citenamefont{Chen, Lewis, Engel,
  Tsui, Ye, Wang, Pfeiffer, and West}}]{Chen04}
\bibinfo{author}{\bibfnamefont{Y.~P.} \bibnamefont{Chen}},
  \bibinfo{author}{\bibfnamefont{R.~M.} \bibnamefont{Lewis}},
  \bibinfo{author}{\bibfnamefont{L.~W.} \bibnamefont{Engel}},
  \bibinfo{author}{\bibfnamefont{D.~C.} \bibnamefont{Tsui}},
  \bibinfo{author}{\bibfnamefont{P.~D.} \bibnamefont{Ye}},
  \bibinfo{author}{\bibfnamefont{Z.~H.} \bibnamefont{Wang}},
  \bibinfo{author}{\bibfnamefont{L.~N.} \bibnamefont{Pfeiffer}},
  \bibnamefont{and} \bibinfo{author}{\bibfnamefont{K.~W.} \bibnamefont{West}},
  \bibinfo{journal}{Phys. Rev. Lett.} \textbf{\bibinfo{volume}{93}},
  \bibinfo{pages}{206805} (\bibinfo{year}{2004}).

\bibitem[{\citenamefont{Cs\'{a}thy et~al.}(2005)\citenamefont{Cs\'{a}thy, Noh,
  Tsui, Pfeiffer, and West}}]{Csathy05}
\bibinfo{author}{\bibfnamefont{G.~A.} \bibnamefont{Cs\'{a}thy}},
  \bibinfo{author}{\bibfnamefont{H.}~\bibnamefont{Noh}},
  \bibinfo{author}{\bibfnamefont{D.~C.} \bibnamefont{Tsui}},
  \bibinfo{author}{\bibfnamefont{L.~N.} \bibnamefont{Pfeiffer}},
  \bibnamefont{and} \bibinfo{author}{\bibfnamefont{K.~W.} \bibnamefont{West}},
  \bibinfo{journal}{Phys. Rev. Lett.} \textbf{\bibinfo{volume}{94}},
  \bibinfo{pages}{226802} (\bibinfo{year}{2005}).

\bibitem[{\citenamefont{Sambandamurthy
  et~al.}(2006)\citenamefont{Sambandamurthy, Wang, Lewis, Chen, Engel, Tsui,
  Pfeiffer, and West}}]{Sambandamurthy06}
\bibinfo{author}{\bibfnamefont{G.}~\bibnamefont{Sambandamurthy}},
  \bibinfo{author}{\bibfnamefont{Z.}~\bibnamefont{Wang}},
  \bibinfo{author}{\bibfnamefont{R.}~\bibnamefont{Lewis}},
  \bibinfo{author}{\bibfnamefont{Y.~P.} \bibnamefont{Chen}},
  \bibinfo{author}{\bibfnamefont{L.}~\bibnamefont{Engel}},
  \bibinfo{author}{\bibfnamefont{D.}~\bibnamefont{Tsui}},
  \bibinfo{author}{\bibfnamefont{L.}~\bibnamefont{Pfeiffer}}, \bibnamefont{and}
  \bibinfo{author}{\bibfnamefont{K.}~\bibnamefont{West}},
  \bibinfo{journal}{Solid State Commun.} \textbf{\bibinfo{volume}{140}},
  \bibinfo{pages}{100 } (\bibinfo{year}{2006}).

\bibitem[{\citenamefont{Chen et~al.}(2006)\citenamefont{Chen, Sambandamurthy,
  Wang, Lewis, Engel, Tsui, Ye, Pfeiffer, and West}}]{Chen06}
\bibinfo{author}{\bibfnamefont{Y.~P.} \bibnamefont{Chen}},
  \bibinfo{author}{\bibfnamefont{G.}~\bibnamefont{Sambandamurthy}},
  \bibinfo{author}{\bibfnamefont{Z.~H.} \bibnamefont{Wang}},
  \bibinfo{author}{\bibfnamefont{R.~M.} \bibnamefont{Lewis}},
  \bibinfo{author}{\bibfnamefont{L.~W.} \bibnamefont{Engel}},
  \bibinfo{author}{\bibfnamefont{D.~C.} \bibnamefont{Tsui}},
  \bibinfo{author}{\bibfnamefont{P.~D.} \bibnamefont{Ye}},
  \bibinfo{author}{\bibfnamefont{L.~N.} \bibnamefont{Pfeiffer}},
  \bibnamefont{and} \bibinfo{author}{\bibfnamefont{K.~W.} \bibnamefont{West}},
  \bibinfo{journal}{Nature Phys.} \textbf{\bibinfo{volume}{2}},
  \bibinfo{pages}{452} (\bibinfo{year}{2006}).

\bibitem[{\citenamefont{Deng et~al.}(2016)\citenamefont{Deng, Liu, Jo,
  Pfeiffer, West, Baldwin, and Shayegan}}]{Deng16}
\bibinfo{author}{\bibfnamefont{H.}~\bibnamefont{Deng}},
  \bibinfo{author}{\bibfnamefont{Y.}~\bibnamefont{Liu}},
  \bibinfo{author}{\bibfnamefont{I.}~\bibnamefont{Jo}},
  \bibinfo{author}{\bibfnamefont{L.~N.} \bibnamefont{Pfeiffer}},
  \bibinfo{author}{\bibfnamefont{K.~W.} \bibnamefont{West}},
  \bibinfo{author}{\bibfnamefont{K.~W.} \bibnamefont{Baldwin}},
  \bibnamefont{and} \bibinfo{author}{\bibfnamefont{M.}~\bibnamefont{Shayegan}},
  \bibinfo{journal}{Phys. Rev. Lett.} \textbf{\bibinfo{volume}{117}},
  \bibinfo{pages}{096601} (\bibinfo{year}{2016}),
  \urlprefix\url{http://link.aps.org/doi/10.1103/PhysRevLett.117.096601}.

\bibitem[{\citenamefont{Yi and Fertig}(1998)}]{Yi98}
\bibinfo{author}{\bibfnamefont{H.}~\bibnamefont{Yi}} \bibnamefont{and}
  \bibinfo{author}{\bibfnamefont{H.~A.} \bibnamefont{Fertig}},
  \bibinfo{journal}{Phys. Rev. B} \textbf{\bibinfo{volume}{58}},
  \bibinfo{pages}{4019} (\bibinfo{year}{1998}).

\bibitem[{\citenamefont{Narevich et~al.}(2001)\citenamefont{Narevich, Murthy,
  and Fertig}}]{Narevich01}
\bibinfo{author}{\bibfnamefont{R.}~\bibnamefont{Narevich}},
  \bibinfo{author}{\bibfnamefont{G.}~\bibnamefont{Murthy}}, \bibnamefont{and}
  \bibinfo{author}{\bibfnamefont{H.~A.} \bibnamefont{Fertig}},
  \bibinfo{journal}{Phys. Rev. B} \textbf{\bibinfo{volume}{64}},
  \bibinfo{pages}{245326} (\bibinfo{year}{2001}).

\bibitem[{\citenamefont{Chang et~al.}(2005)\citenamefont{Chang, Jeon, and
  Jain}}]{Chang05}
\bibinfo{author}{\bibfnamefont{C.-C.} \bibnamefont{Chang}},
  \bibinfo{author}{\bibfnamefont{G.~S.} \bibnamefont{Jeon}}, \bibnamefont{and}
  \bibinfo{author}{\bibfnamefont{J.~K.} \bibnamefont{Jain}},
  \bibinfo{journal}{Phys. Rev. Lett.} \textbf{\bibinfo{volume}{94}},
  \bibinfo{pages}{016809} (\bibinfo{year}{2005}).

\bibitem[{\citenamefont{Archer et~al.}(2013)\citenamefont{Archer, Park, and
  Jain}}]{Archer13}
\bibinfo{author}{\bibfnamefont{A.~C.} \bibnamefont{Archer}},
  \bibinfo{author}{\bibfnamefont{K.}~\bibnamefont{Park}}, \bibnamefont{and}
  \bibinfo{author}{\bibfnamefont{J.~K.} \bibnamefont{Jain}},
  \bibinfo{journal}{Phys. Rev. Lett.} \textbf{\bibinfo{volume}{111}},
  \bibinfo{pages}{146804} (\bibinfo{year}{2013}).

\bibitem[{\citenamefont{Maki and Zotos}(1983)}]{Maki83}
\bibinfo{author}{\bibfnamefont{K.}~\bibnamefont{Maki}} \bibnamefont{and}
  \bibinfo{author}{\bibfnamefont{X.}~\bibnamefont{Zotos}},
  \bibinfo{journal}{Phys. Rev. B} \textbf{\bibinfo{volume}{28}},
  \bibinfo{pages}{4349} (\bibinfo{year}{1983}).

\bibitem[{\citenamefont{Liu et~al.}(2014)\citenamefont{Liu, Kamburov, Hasdemir,
  Shayegan, Pfeiffer, West, and Baldwin}}]{Liu14a}
\bibinfo{author}{\bibfnamefont{Y.}~\bibnamefont{Liu}},
  \bibinfo{author}{\bibfnamefont{D.}~\bibnamefont{Kamburov}},
  \bibinfo{author}{\bibfnamefont{S.}~\bibnamefont{Hasdemir}},
  \bibinfo{author}{\bibfnamefont{M.}~\bibnamefont{Shayegan}},
  \bibinfo{author}{\bibfnamefont{L.~N.} \bibnamefont{Pfeiffer}},
  \bibinfo{author}{\bibfnamefont{K.~W.} \bibnamefont{West}}, \bibnamefont{and}
  \bibinfo{author}{\bibfnamefont{K.~W.} \bibnamefont{Baldwin}},
  \bibinfo{journal}{Phys. Rev. Lett.} \textbf{\bibinfo{volume}{113}},
  \bibinfo{pages}{246803} (\bibinfo{year}{2014}),
  \urlprefix\url{http://link.aps.org/doi/10.1103/PhysRevLett.113.246803}.

\bibitem[{\citenamefont{Zhang et~al.}(2015)\citenamefont{Zhang, Du, Manfra,
  Pfeiffer, and West}}]{Zhang15c}
\bibinfo{author}{\bibfnamefont{C.}~\bibnamefont{Zhang}},
  \bibinfo{author}{\bibfnamefont{R.-R.} \bibnamefont{Du}},
  \bibinfo{author}{\bibfnamefont{M.~J.} \bibnamefont{Manfra}},
  \bibinfo{author}{\bibfnamefont{L.~N.} \bibnamefont{Pfeiffer}},
  \bibnamefont{and} \bibinfo{author}{\bibfnamefont{K.~W.} \bibnamefont{West}},
  \bibinfo{journal}{Phys. Rev. B} \textbf{\bibinfo{volume}{92}},
  \bibinfo{pages}{075434} (\bibinfo{year}{2015}),
  \urlprefix\url{https://link.aps.org/doi/10.1103/PhysRevB.92.075434}.

\bibitem[{\citenamefont{Jang et~al.}(2017)\citenamefont{Jang, Hunt, Pfeiffer,
  West, and Ashoori}}]{Jang17}
\bibinfo{author}{\bibfnamefont{J.}~\bibnamefont{Jang}},
  \bibinfo{author}{\bibfnamefont{B.~M.} \bibnamefont{Hunt}},
  \bibinfo{author}{\bibfnamefont{L.~N.} \bibnamefont{Pfeiffer}},
  \bibinfo{author}{\bibfnamefont{K.~W.} \bibnamefont{West}}, \bibnamefont{and}
  \bibinfo{author}{\bibfnamefont{R.~C.} \bibnamefont{Ashoori}},
  \bibinfo{journal}{Nature Physics} \textbf{\bibinfo{volume}{13}},
  \bibinfo{pages}{340} (\bibinfo{year}{2017}), ISSN \bibinfo{issn}{1745-2473}.

\bibitem[{\citenamefont{Santos et~al.}(1992{\natexlab{a}})\citenamefont{Santos,
  Suen, Shayegan, Li, Engel, and Tsui}}]{Santos92}
\bibinfo{author}{\bibfnamefont{M.~B.} \bibnamefont{Santos}},
  \bibinfo{author}{\bibfnamefont{Y.~W.} \bibnamefont{Suen}},
  \bibinfo{author}{\bibfnamefont{M.}~\bibnamefont{Shayegan}},
  \bibinfo{author}{\bibfnamefont{Y.~P.} \bibnamefont{Li}},
  \bibinfo{author}{\bibfnamefont{L.~W.} \bibnamefont{Engel}}, \bibnamefont{and}
  \bibinfo{author}{\bibfnamefont{D.~C.} \bibnamefont{Tsui}},
  \bibinfo{journal}{Phys. Rev. Lett.} \textbf{\bibinfo{volume}{68}},
  \bibinfo{pages}{1188} (\bibinfo{year}{1992}{\natexlab{a}}).

\bibitem[{\citenamefont{Santos et~al.}(1992{\natexlab{b}})\citenamefont{Santos,
  Jo, Suen, Engel, and Shayegan}}]{Santos92b}
\bibinfo{author}{\bibfnamefont{M.~B.} \bibnamefont{Santos}},
  \bibinfo{author}{\bibfnamefont{J.}~\bibnamefont{Jo}},
  \bibinfo{author}{\bibfnamefont{Y.~W.} \bibnamefont{Suen}},
  \bibinfo{author}{\bibfnamefont{L.~W.} \bibnamefont{Engel}}, \bibnamefont{and}
  \bibinfo{author}{\bibfnamefont{M.}~\bibnamefont{Shayegan}},
  \bibinfo{journal}{Phys. Rev. B} \textbf{\bibinfo{volume}{46}},
  \bibinfo{pages}{13639} (\bibinfo{year}{1992}{\natexlab{b}}),
  \urlprefix\url{https://link.aps.org/doi/10.1103/PhysRevB.46.13639}.

\bibitem[{\citenamefont{Pan et~al.}(2005)\citenamefont{Pan, Cs\'athy, Tsui,
  Pfeiffer, and West}}]{Pan05}
\bibinfo{author}{\bibfnamefont{W.}~\bibnamefont{Pan}},
  \bibinfo{author}{\bibfnamefont{G.~A.} \bibnamefont{Cs\'athy}},
  \bibinfo{author}{\bibfnamefont{D.~C.} \bibnamefont{Tsui}},
  \bibinfo{author}{\bibfnamefont{L.~N.} \bibnamefont{Pfeiffer}},
  \bibnamefont{and} \bibinfo{author}{\bibfnamefont{K.~W.} \bibnamefont{West}},
  \bibinfo{journal}{Phys. Rev. B} \textbf{\bibinfo{volume}{71}},
  \bibinfo{pages}{035302} (\bibinfo{year}{2005}),
  \urlprefix\url{https://link.aps.org/doi/10.1103/PhysRevB.71.035302}.

\bibitem[{\citenamefont{Zhu and Louie}(1993)}]{Zhu93}
\bibinfo{author}{\bibfnamefont{X.}~\bibnamefont{Zhu}} \bibnamefont{and}
  \bibinfo{author}{\bibfnamefont{S.~G.} \bibnamefont{Louie}},
  \bibinfo{journal}{Phys. Rev. Lett.} \textbf{\bibinfo{volume}{70}},
  \bibinfo{pages}{335} (\bibinfo{year}{1993}).

\bibitem[{\citenamefont{Price et~al.}(1993)\citenamefont{Price, Platzman, and
  He}}]{Price93}
\bibinfo{author}{\bibfnamefont{R.}~\bibnamefont{Price}},
  \bibinfo{author}{\bibfnamefont{P.~M.} \bibnamefont{Platzman}},
  \bibnamefont{and} \bibinfo{author}{\bibfnamefont{S.}~\bibnamefont{He}},
  \bibinfo{journal}{Phys. Rev. Lett.} \textbf{\bibinfo{volume}{70}},
  \bibinfo{pages}{339} (\bibinfo{year}{1993}).

\bibitem[{\citenamefont{Platzman and Price}(1993)}]{Platzman93}
\bibinfo{author}{\bibfnamefont{P.~M.} \bibnamefont{Platzman}} \bibnamefont{and}
  \bibinfo{author}{\bibfnamefont{R.}~\bibnamefont{Price}},
  \bibinfo{journal}{Phys. Rev. Lett.} \textbf{\bibinfo{volume}{70}},
  \bibinfo{pages}{3487} (\bibinfo{year}{1993}).

\bibitem[{\citenamefont{Ortiz et~al.}(1993)\citenamefont{Ortiz, Ceperley, and
  Martin}}]{Ortiz93}
\bibinfo{author}{\bibfnamefont{G.}~\bibnamefont{Ortiz}},
  \bibinfo{author}{\bibfnamefont{D.~M.} \bibnamefont{Ceperley}},
  \bibnamefont{and} \bibinfo{author}{\bibfnamefont{R.~M.}
  \bibnamefont{Martin}}, \bibinfo{journal}{Phys. Rev. Lett.}
  \textbf{\bibinfo{volume}{71}}, \bibinfo{pages}{2777} (\bibinfo{year}{1993}),
  \urlprefix\url{http://link.aps.org/doi/10.1103/PhysRevLett.71.2777}.

\bibitem[{\citenamefont{He et~al.}(2005)\citenamefont{He, Cui, Ma, Chen, Liu,
  and Zou}}]{He05}
\bibinfo{author}{\bibfnamefont{W.~J.} \bibnamefont{He}},
  \bibinfo{author}{\bibfnamefont{T.}~\bibnamefont{Cui}},
  \bibinfo{author}{\bibfnamefont{Y.~M.} \bibnamefont{Ma}},
  \bibinfo{author}{\bibfnamefont{C.~B.} \bibnamefont{Chen}},
  \bibinfo{author}{\bibfnamefont{Z.~M.} \bibnamefont{Liu}}, \bibnamefont{and}
  \bibinfo{author}{\bibfnamefont{G.~T.} \bibnamefont{Zou}},
  \bibinfo{journal}{Phys. Rev. B} \textbf{\bibinfo{volume}{72}},
  \bibinfo{pages}{195306} (\bibinfo{year}{2005}).

\bibitem[{\citenamefont{Maryenko et~al.}(2017)\citenamefont{Maryenko, McCollam,
  Falson, Kozuka, Bruin, Zeitler, and Kawasaki}}]{Maryenko17}
\bibinfo{author}{\bibfnamefont{D.}~\bibnamefont{Maryenko}},
  \bibinfo{author}{\bibfnamefont{A.}~\bibnamefont{McCollam}},
  \bibinfo{author}{\bibfnamefont{J.}~\bibnamefont{Falson}},
  \bibinfo{author}{\bibfnamefont{Y.}~\bibnamefont{Kozuka}},
  \bibinfo{author}{\bibfnamefont{J.}~\bibnamefont{Bruin}},
  \bibinfo{author}{\bibfnamefont{U.}~\bibnamefont{Zeitler}}, \bibnamefont{and}
  \bibinfo{author}{\bibfnamefont{M.}~\bibnamefont{Kawasaki}}
  (\bibinfo{year}{2017}), \urlprefix\url{arXiv:1707.08406}.

\bibitem[{SM-()}]{SM-Zhao-2018}
\bibinfo{note}{See Supplemental Material which includes a brief review of the
  fixed-phase DMC calculation; details of thermodynamic extrapolation;
  comparison between the $\kappa=0$ energies of the electron and $^2$CF
  crystals at $\nu=1/5$ and 2/9; and finite width results.}

\bibitem[{\citenamefont{Melton and Mitas}(2017)}]{Melton17}
\bibinfo{author}{\bibfnamefont{C.~A.} \bibnamefont{Melton}} \bibnamefont{and}
  \bibinfo{author}{\bibfnamefont{L.}~\bibnamefont{Mitas}},
  \bibinfo{journal}{Phys. Rev. E} \textbf{\bibinfo{volume}{96}},
  \bibinfo{pages}{043305} (\bibinfo{year}{2017}),
  \urlprefix\url{https://link.aps.org/doi/10.1103/PhysRevE.96.043305}.

\bibitem[{\citenamefont{Reynolds et~al.}(1982)\citenamefont{Reynolds, Ceperley,
  Alder, and Lester~Jr.}}]{Reynolds82}
\bibinfo{author}{\bibfnamefont{P.~J.} \bibnamefont{Reynolds}},
  \bibinfo{author}{\bibfnamefont{D.~M.} \bibnamefont{Ceperley}},
  \bibinfo{author}{\bibfnamefont{B.~J.} \bibnamefont{Alder}}, \bibnamefont{and}
  \bibinfo{author}{\bibfnamefont{W.~A.} \bibnamefont{Lester~Jr.}},
  \bibinfo{journal}{J. Chem. Phys.} \textbf{\bibinfo{volume}{77}},
  \bibinfo{pages}{5593} (\bibinfo{year}{1982}),
  \urlprefix\url{http://scitation.aip.org/content/aip/journal/jcp/77/11/10.106%
3/1.443766}.

\bibitem[{\citenamefont{Foulkes et~al.}(2001)\citenamefont{Foulkes, Mitas,
  Needs, and Rajagopal}}]{Foulkes01}
\bibinfo{author}{\bibfnamefont{W.~M.~C.} \bibnamefont{Foulkes}},
  \bibinfo{author}{\bibfnamefont{L.}~\bibnamefont{Mitas}},
  \bibinfo{author}{\bibfnamefont{R.~J.} \bibnamefont{Needs}}, \bibnamefont{and}
  \bibinfo{author}{\bibfnamefont{G.}~\bibnamefont{Rajagopal}},
  \bibinfo{journal}{Rev. Mod. Phys.} \textbf{\bibinfo{volume}{73}},
  \bibinfo{pages}{33} (\bibinfo{year}{2001}),
  \urlprefix\url{http://link.aps.org/doi/10.1103/RevModPhys.73.33}.

\bibitem[{\citenamefont{Haldane}(1983)}]{Haldane83}
\bibinfo{author}{\bibfnamefont{F.~D.~M.} \bibnamefont{Haldane}},
  \bibinfo{journal}{Phys. Rev. Lett.} \textbf{\bibinfo{volume}{51}},
  \bibinfo{pages}{605} (\bibinfo{year}{1983}),
  \urlprefix\url{http://link.aps.org/doi/10.1103/PhysRevLett.51.605}.

\bibitem[{\citenamefont{Melik-Alaverdian
  et~al.}(1997)\citenamefont{Melik-Alaverdian, Bonesteel, and
  Ortiz}}]{Melik-Alaverdian97}
\bibinfo{author}{\bibfnamefont{V.}~\bibnamefont{Melik-Alaverdian}},
  \bibinfo{author}{\bibfnamefont{N.~E.} \bibnamefont{Bonesteel}},
  \bibnamefont{and} \bibinfo{author}{\bibfnamefont{G.}~\bibnamefont{Ortiz}},
  \bibinfo{journal}{Phys. Rev. Lett.} \textbf{\bibinfo{volume}{79}},
  \bibinfo{pages}{5286} (\bibinfo{year}{1997}),
  \urlprefix\url{http://link.aps.org/doi/10.1103/PhysRevLett.79.5286}.

\bibitem[{\citenamefont{G\"u\c{c}l\"u and Umrigar}(2005)}]{Guclu05}
\bibinfo{author}{\bibfnamefont{A.~D.} \bibnamefont{G\"u\c{c}l\"u}}
  \bibnamefont{and} \bibinfo{author}{\bibfnamefont{C.~J.}
  \bibnamefont{Umrigar}}, \bibinfo{journal}{Phys. Rev. B}
  \textbf{\bibinfo{volume}{72}}, \bibinfo{pages}{045309}
  (\bibinfo{year}{2005}),
  \urlprefix\url{http://link.aps.org/doi/10.1103/PhysRevB.72.045309}.

\bibitem[{\citenamefont{Zhang et~al.}(2016)\citenamefont{Zhang, W\'ojs, and
  Jain}}]{Zhang16}
\bibinfo{author}{\bibfnamefont{Y.}~\bibnamefont{Zhang}},
  \bibinfo{author}{\bibfnamefont{A.}~\bibnamefont{W\'ojs}}, \bibnamefont{and}
  \bibinfo{author}{\bibfnamefont{J.~K.} \bibnamefont{Jain}},
  \bibinfo{journal}{Phys. Rev. Lett.} \textbf{\bibinfo{volume}{117}},
  \bibinfo{pages}{116803} (\bibinfo{year}{2016}),
  \urlprefix\url{http://link.aps.org/doi/10.1103/PhysRevLett.117.116803}.

\bibitem[{\citenamefont{Zhang et~al.}(2017)\citenamefont{Zhang, Jain, and
  Eisenstein}}]{Zhang17}
\bibinfo{author}{\bibfnamefont{Y.}~\bibnamefont{Zhang}},
  \bibinfo{author}{\bibfnamefont{J.~K.} \bibnamefont{Jain}}, \bibnamefont{and}
  \bibinfo{author}{\bibfnamefont{J.~P.} \bibnamefont{Eisenstein}},
  \bibinfo{journal}{Phys. Rev. B} \textbf{\bibinfo{volume}{95}},
  \bibinfo{pages}{195105} (\bibinfo{year}{2017}),
  \urlprefix\url{https://link.aps.org/doi/10.1103/PhysRevB.95.195105}.

\bibitem[{\citenamefont{Thomson}(1904)}]{Thomson04}
\bibinfo{author}{\bibfnamefont{J.~J.} \bibnamefont{Thomson}},
  \bibinfo{journal}{Phil. Mag.} \textbf{\bibinfo{volume}{7}},
  \bibinfo{pages}{237} (\bibinfo{year}{1904}).

\bibitem[{\citenamefont{Wales and Ulker}(2006)}]{Wales06}
\bibinfo{author}{\bibfnamefont{D.~J.} \bibnamefont{Wales}} \bibnamefont{and}
  \bibinfo{author}{\bibfnamefont{S.}~\bibnamefont{Ulker}},
  \bibinfo{journal}{Phys. Rev. B} \textbf{\bibinfo{volume}{74}},
  \bibinfo{pages}{212101} (\bibinfo{year}{2006}),
  \urlprefix\url{https://link.aps.org/doi/10.1103/PhysRevB.74.212101}.

\bibitem[{\citenamefont{Wales et~al.}(2009)\citenamefont{Wales, McKay, and
  Altschuler}}]{Wales09}
\bibinfo{author}{\bibfnamefont{D.~J.} \bibnamefont{Wales}},
  \bibinfo{author}{\bibfnamefont{H.}~\bibnamefont{McKay}}, \bibnamefont{and}
  \bibinfo{author}{\bibfnamefont{E.~L.} \bibnamefont{Altschuler}},
  \bibinfo{journal}{Phys. Rev. B} \textbf{\bibinfo{volume}{79}},
  \bibinfo{pages}{224115} (\bibinfo{year}{2009}),
  \urlprefix\url{https://link.aps.org/doi/10.1103/PhysRevB.79.224115}.

\bibitem[{Tho()}]{Thomson}
\bibinfo{note}{The minimum energy locations can be found at
  \url{http://thomson.phy.syr.edu/}}.

\bibitem[{\citenamefont{Jain and Kamilla}(1997{\natexlab{a}})}]{Jain97}
\bibinfo{author}{\bibfnamefont{J.~K.} \bibnamefont{Jain}} \bibnamefont{and}
  \bibinfo{author}{\bibfnamefont{R.~K.} \bibnamefont{Kamilla}},
  \bibinfo{journal}{Int. J. Mod. Phys. B} \textbf{\bibinfo{volume}{11}},
  \bibinfo{pages}{2621} (\bibinfo{year}{1997}{\natexlab{a}}).

\bibitem[{\citenamefont{Jain and Kamilla}(1997{\natexlab{b}})}]{Jain97b}
\bibinfo{author}{\bibfnamefont{J.~K.} \bibnamefont{Jain}} \bibnamefont{and}
  \bibinfo{author}{\bibfnamefont{R.~K.} \bibnamefont{Kamilla}},
  \bibinfo{journal}{Phys. Rev. B} \textbf{\bibinfo{volume}{55}},
  \bibinfo{pages}{R4895} (\bibinfo{year}{1997}{\natexlab{b}}),
  \urlprefix\url{http://link.aps.org/doi/10.1103/PhysRevB.55.R4895}.
  
\bibitem[{\citenamefont{Sodemann and MacDonald}(2013)}]{Sodemann13}
\bibinfo{author}{\bibfnamefont{I.}~\bibnamefont{Sodemann}} \bibnamefont{and}
  \bibinfo{author}{\bibfnamefont{A.~H.} \bibnamefont{MacDonald}},
  \bibinfo{journal}{Phys. Rev. B} \textbf{\bibinfo{volume}{87}},
  \bibinfo{pages}{245425} (\bibinfo{year}{2013}),
  \urlprefix\url{http://link.aps.org/doi/10.1103/PhysRevB.87.245425}.

\bibitem[{\citenamefont{Manoharan et~al.}(1996)\citenamefont{Manoharan, Suen,
  Santos, and Shayegan}}]{Manoharan96}
\bibinfo{author}{\bibfnamefont{H.~C.} \bibnamefont{Manoharan}},
  \bibinfo{author}{\bibfnamefont{Y.~W.} \bibnamefont{Suen}},
  \bibinfo{author}{\bibfnamefont{M.~B.} \bibnamefont{Santos}},
  \bibnamefont{and} \bibinfo{author}{\bibfnamefont{M.}~\bibnamefont{Shayegan}},
  \bibinfo{journal}{Phys. Rev. Lett.} \textbf{\bibinfo{volume}{77}},
  \bibinfo{pages}{1813} (\bibinfo{year}{1996}),
  \urlprefix\url{https://link.aps.org/doi/10.1103/PhysRevLett.77.1813}.

\end{thebibliography}
%\bibliographystyle{apsrev}

\pagebreak

\begin{widetext}

\centerline{\bf Supplemental Material for}
\centerline{\bf``Landau-level-mixing induced crystallization in the fractional quantum Hall regime"}

\hspace{2cm}

\setcounter{figure}{0}
\setcounter{equation}{0}
\renewcommand\thefigure{S\arabic{figure}}
\renewcommand\thetable{S\arabic{table}}
\renewcommand\theequation{S\arabic{equation}}

Figs.~S1-S4 show thermodynamic extrapolations for various energies at certain selected filling factors. (The analysis for other filling factors is very similar.) For the liquid energies at zero width, it is possible to obtain the thermodynamic energies by a direct extrapolation of the finite $N$ results, as seen in the top left panels of Figs.~S1-S4. This is not the case for liquids in quantum wells of finite width, and crystals in either zero of finite width quantum wells. For these, we obtain thermodynamic extrapolations separately for (i) $E_{\kappa=0}$, where large systems are accessible; and (ii) $\Delta E_\kappa=E_\kappa-E_{\kappa=0}$ which provides a reliable linear extrapolation even with $N\lesssim 35$ particles.  

Figs.~S5 and S6 show the thermodynamic energies as a function of $\kappa$ for several different filling factors of interest. Fig.~S5 reveals that while the liquid remains the ground state at 1/3 and 2/5 all the way up to $\kappa=20$, a level crossing transition takes place at nearby fillings, such as $\nu=0.320$ and $\nu=0.373$.  Fig.~S6 also compares the energies of the electron crystal and the $^2$CF crystal at $\nu=1/5$ and $\nu=2/9$. 

All figures also show results for three different widths: $w=0$, $w=2l$ and $w=4l$, where $l$ is the magnetic length. As noted in the main text, we assume a quantum well of width $w$ with a transverse wave function $\xi(\eta)=\sqrt{2 /w}\sin{\pi \eta/ w}$, where the transverse coordinate $0<\eta<w$.  The effective two-dimensional interaction is then given by $V^{\rm eff}(r)=\int_0^w d\eta_1\int_0^w d\eta_2 {|\xi(\eta_1)|^2 |\xi(\eta_2)|^2\over [r^2+(\eta_1-\eta_2)^2]^{1/2}}$ where $r=\sqrt{(x_1-x_2)^2+(y_1-y_2)^2}$. 

Fig.~S7 shows how finite width modifies the phase diagram in the vicinity of $\nu=1/3$. It has not been possible to obtain the finite width phase diagram reliably in the vicinity of $\nu=1/5$, because, as seen in Fig.~S6, the energies of the liquid and the crystal remain very close for a large range of $\kappa$.

\begin{figure}
\includegraphics[scale=0.3]{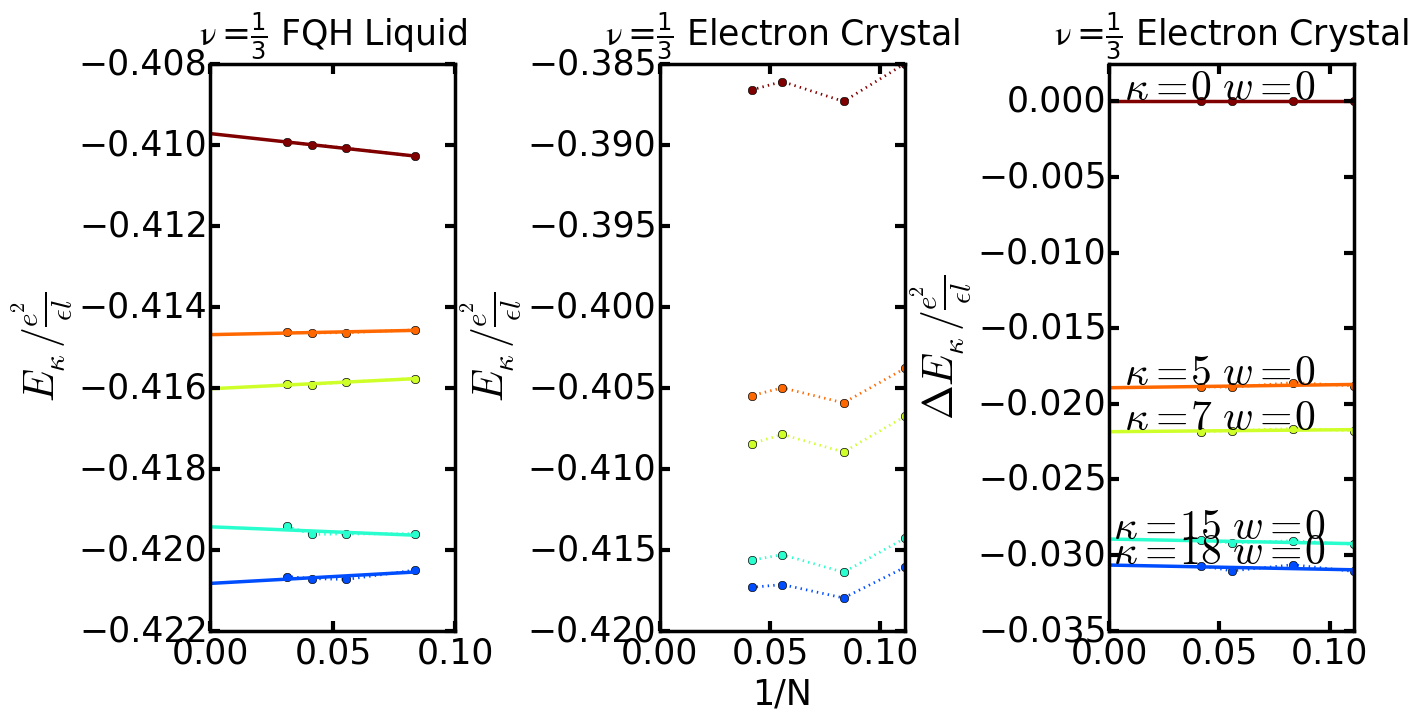}
\includegraphics[scale=0.3]{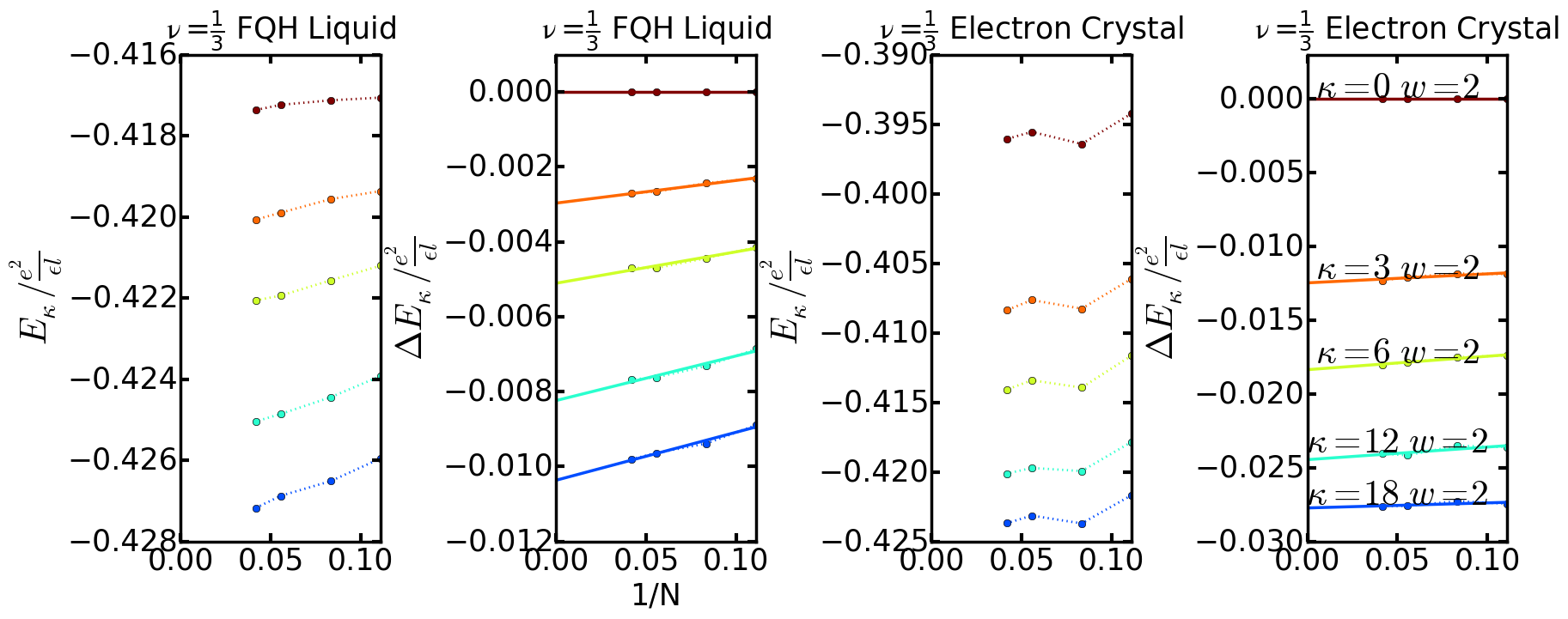}
\includegraphics[scale=0.3]{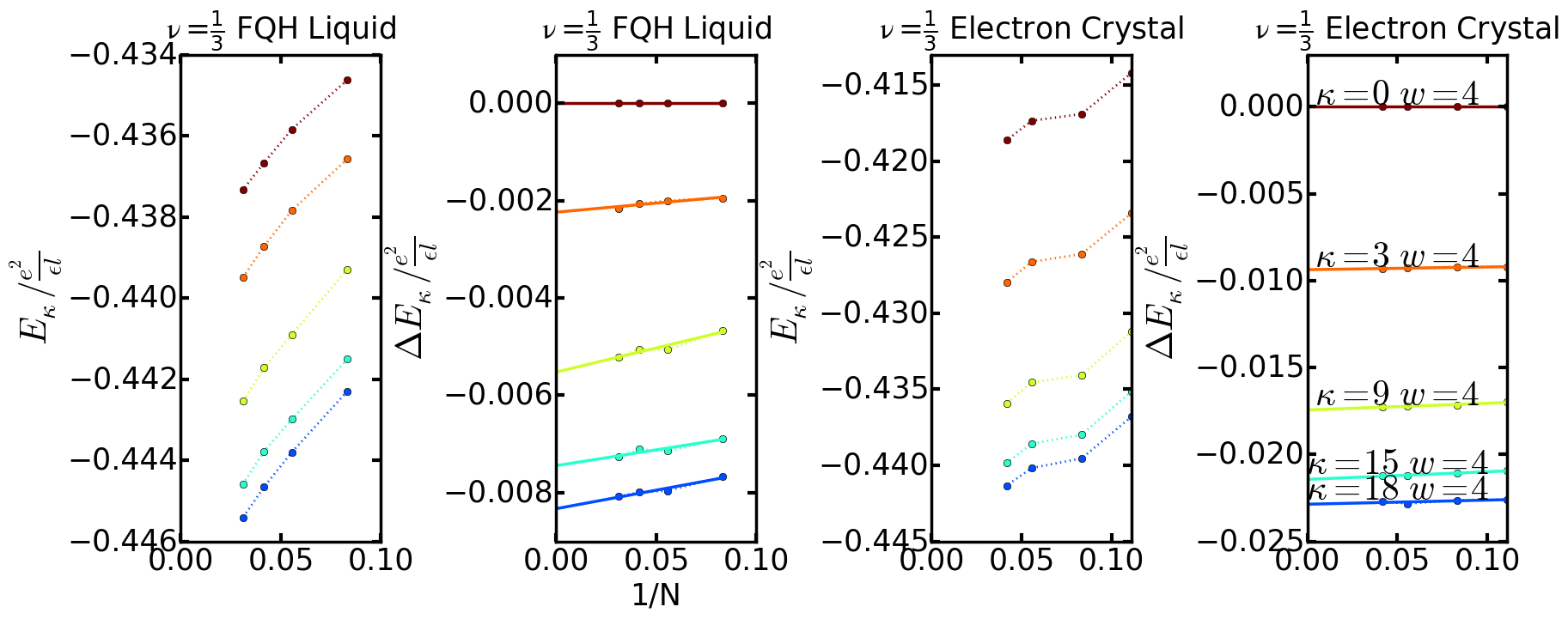}
\includegraphics[scale=0.3]{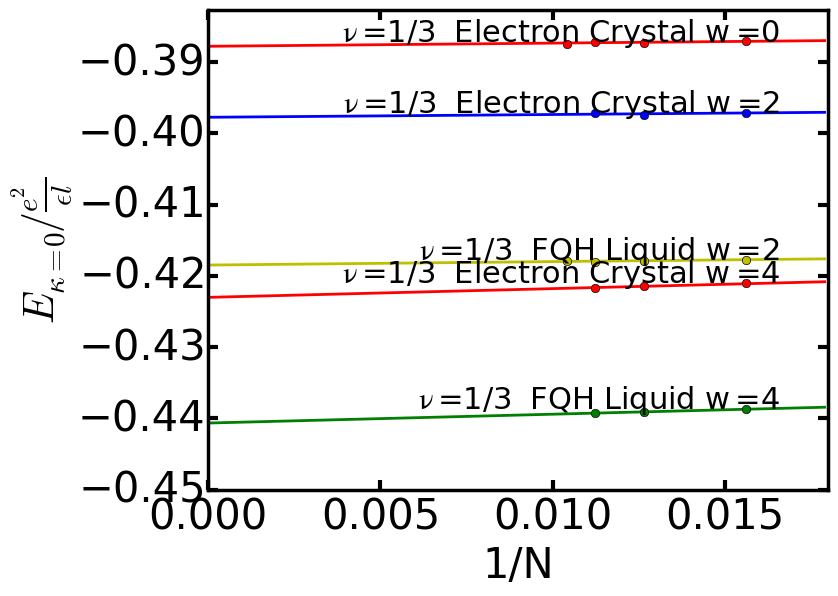}
\caption{
This figure shows extrapolations of the energies of the liquid and crystal states at filling factor $\nu={1\over3}$. 
The dots mark the calculated energies, the dotted lines connect successive dots, and the solid lines represent extrapolations to the thermodynamic limit. The first, second and third rows show results for quantum well widths $w=$0, 2$l$, and 4$l$, respectively, where $l$ is the magnetic length. 
Different colors represent different Landau level mixing $\kappa$ shown in the rightmost plot of each row. $E_\kappa$ is th e energy per particle obtained from the diffusion Monte Carlo method, including interaction with the background. The energy difference $\Delta E_{\kappa}^N$ is defined as $\Delta E_{\kappa}^N=E_{\kappa}^N-E_{\kappa=0}^N$.
For $w=0$, the energy $E_\kappa$ of the liquid state shows a good linear extrapolation as a function of $1/N$ (topmost row). For finite widths $E_\kappa$ bends downward with increasing $N$, making a direct linear extrapolation unsuitable, and therefore we extrapolate the energy difference $\Delta E_{\kappa}$. For the crystal state, $E_\kappa$ has finite size fluctuations due to defects, but $\Delta E_{\kappa}^N$ extrapolates nicely to the thermodynamic limit. 
The last row shows the thermodynamic values for various liquid and crystal energies $E_{\kappa=0}$; here variational Monte Carlo method can access very large systems where finite-size fluctuations are negligible.
}
\label{1_3}
\end{figure}

\begin{figure}
\includegraphics[scale=0.3]{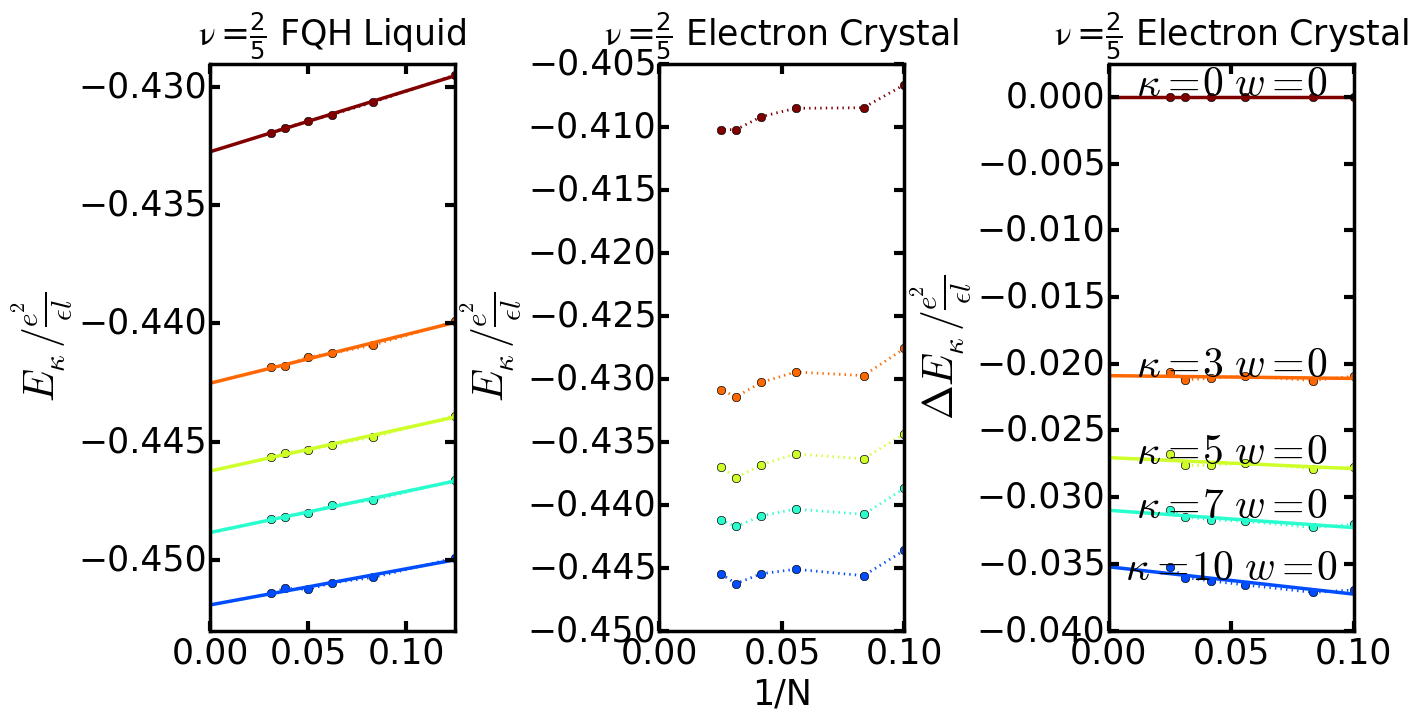}
\includegraphics[scale=0.3]{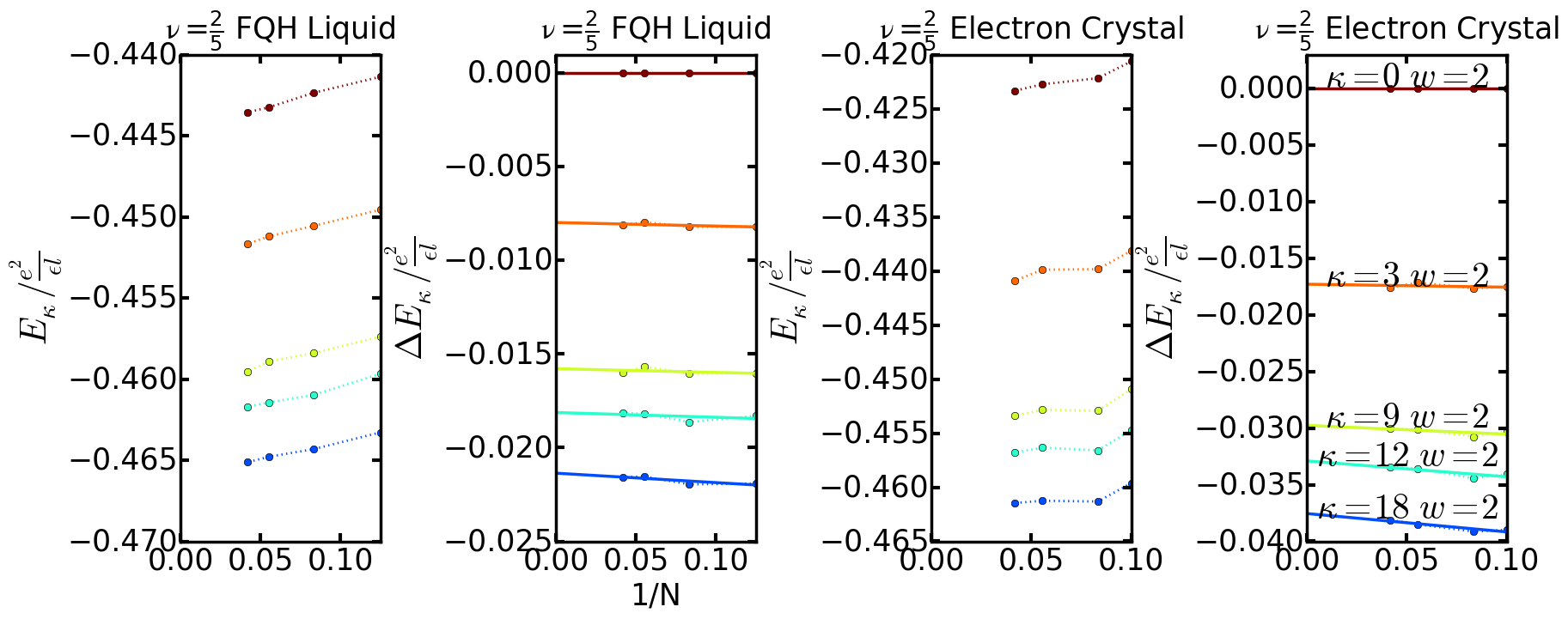}
\includegraphics[scale=0.3]{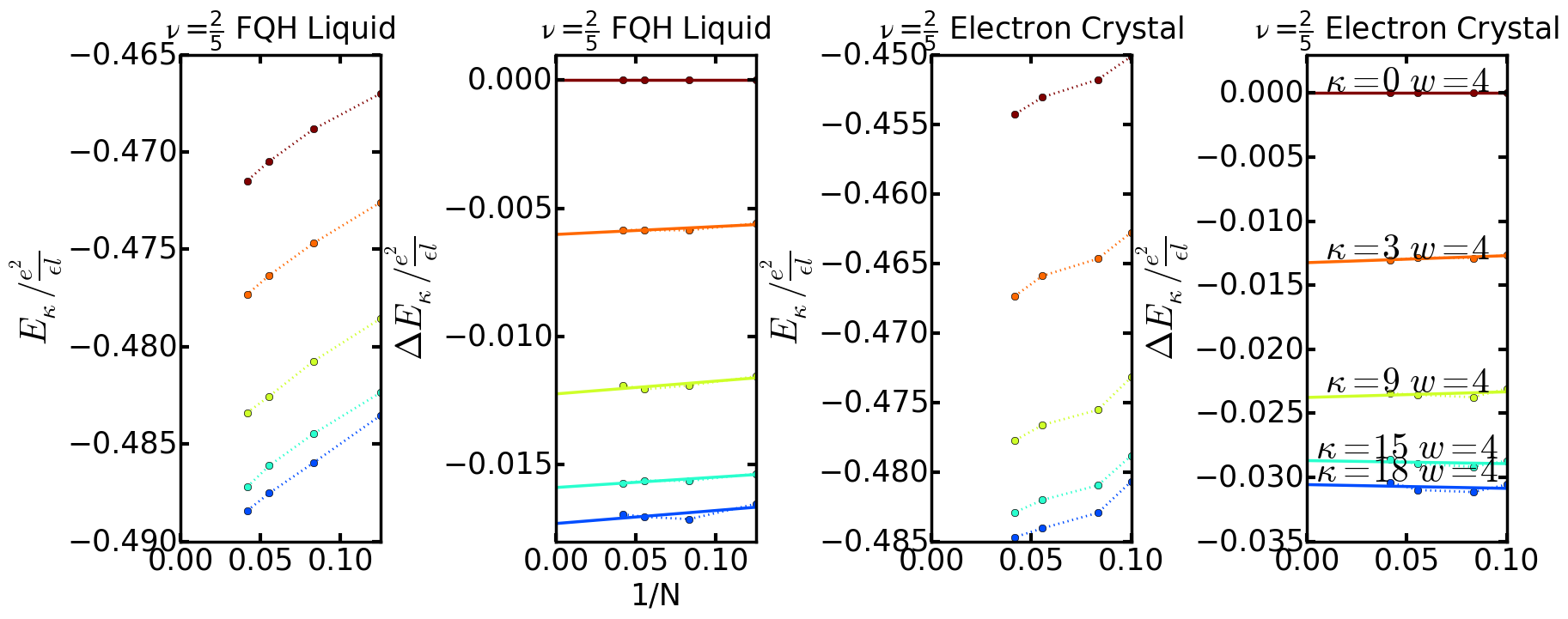}
\includegraphics[scale=0.3]{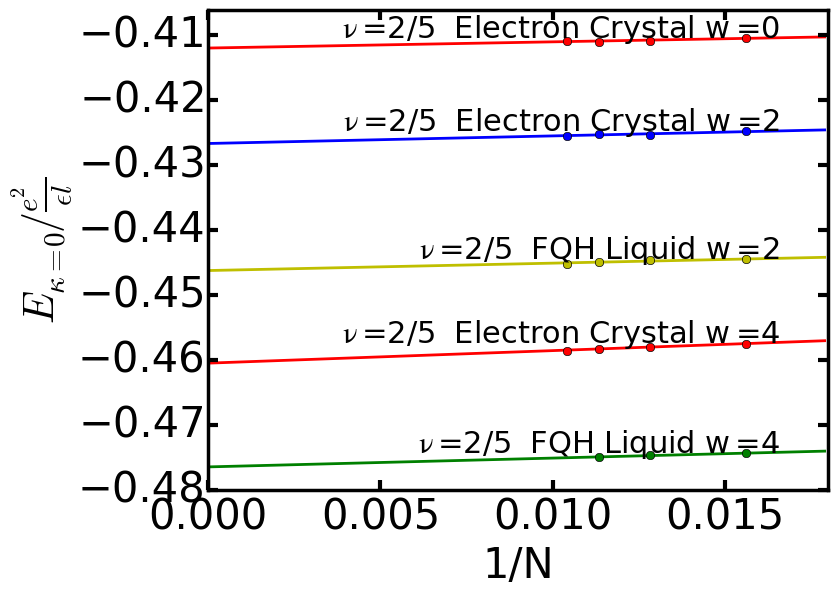}
\caption{
Same as in Fig.~\ref{1_3} but for $\nu=2/5$.
}
\label{2_5}
\end{figure}

\begin{figure}
\includegraphics[scale=0.3]{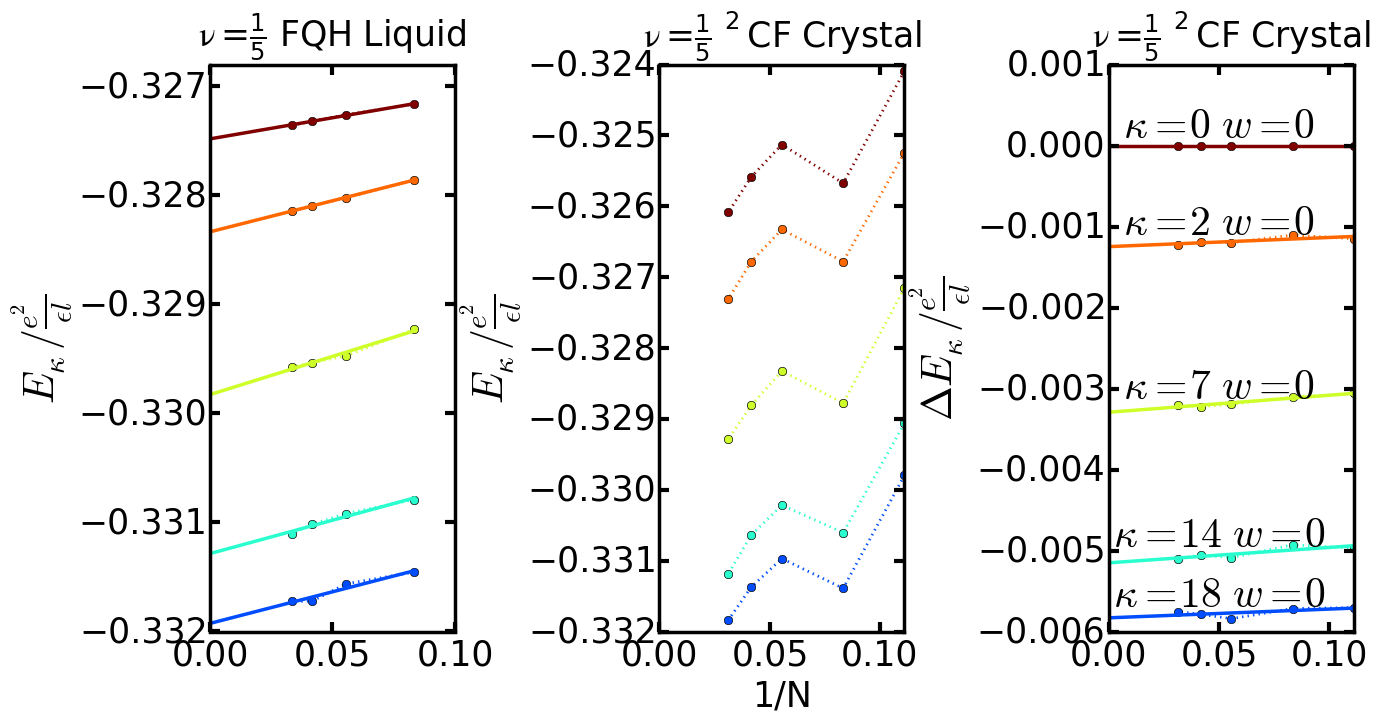}
\includegraphics[scale=0.3]{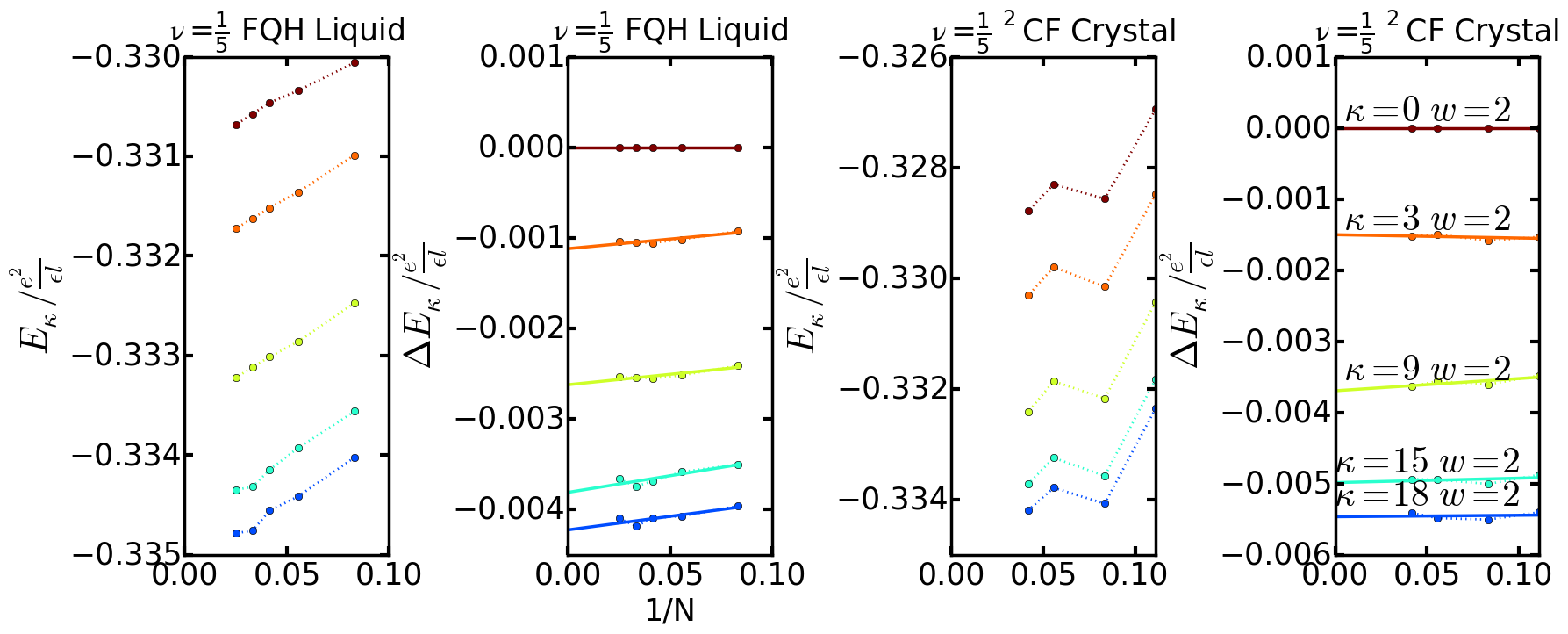}
\includegraphics[scale=0.3]{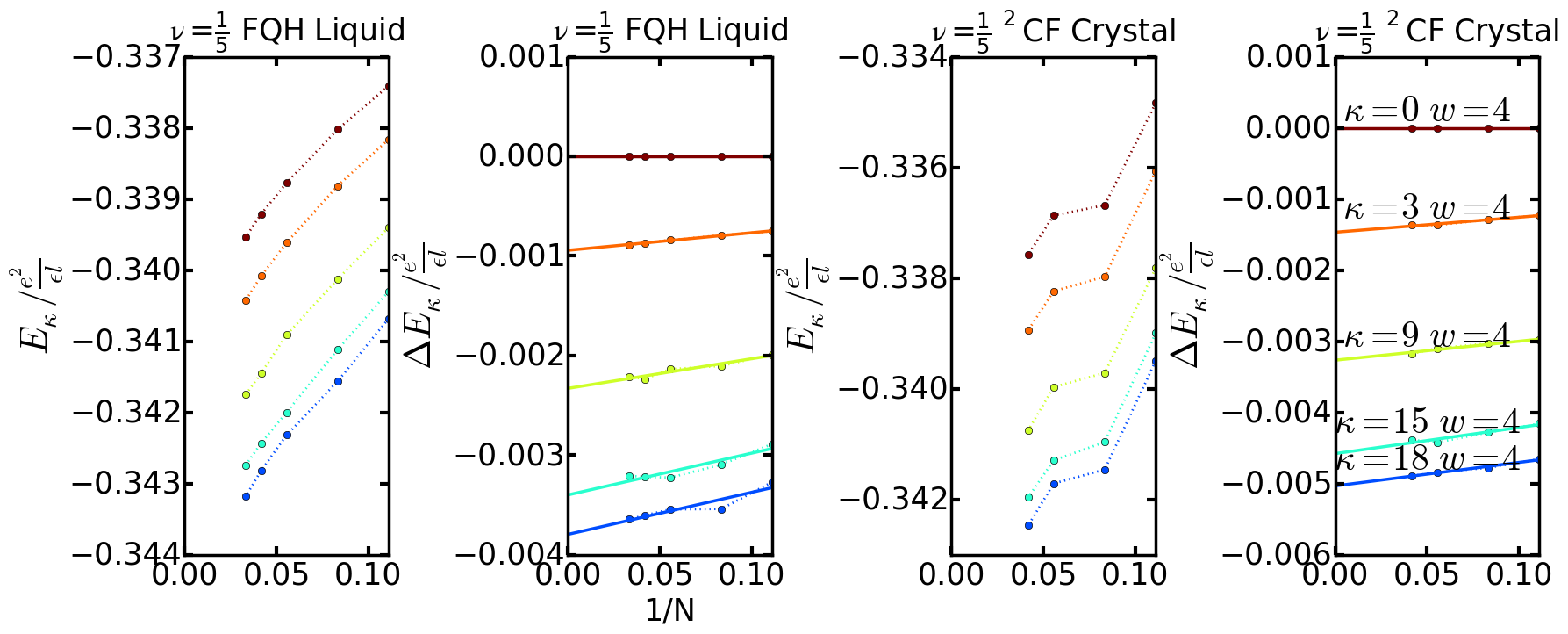}
\includegraphics[scale=0.3]{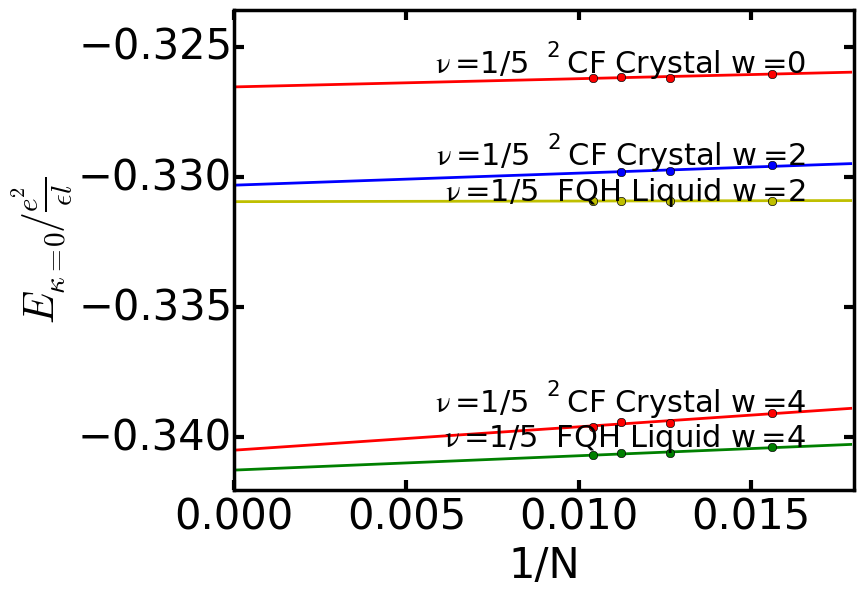}
\caption{
Same as in Fig.~\ref{1_3} but for $\nu=1/5$. Here the relevant crystal is the $^2$CF crystal.
}
\label{1_5}
\end{figure}

\begin{figure}
\includegraphics[scale=0.3]{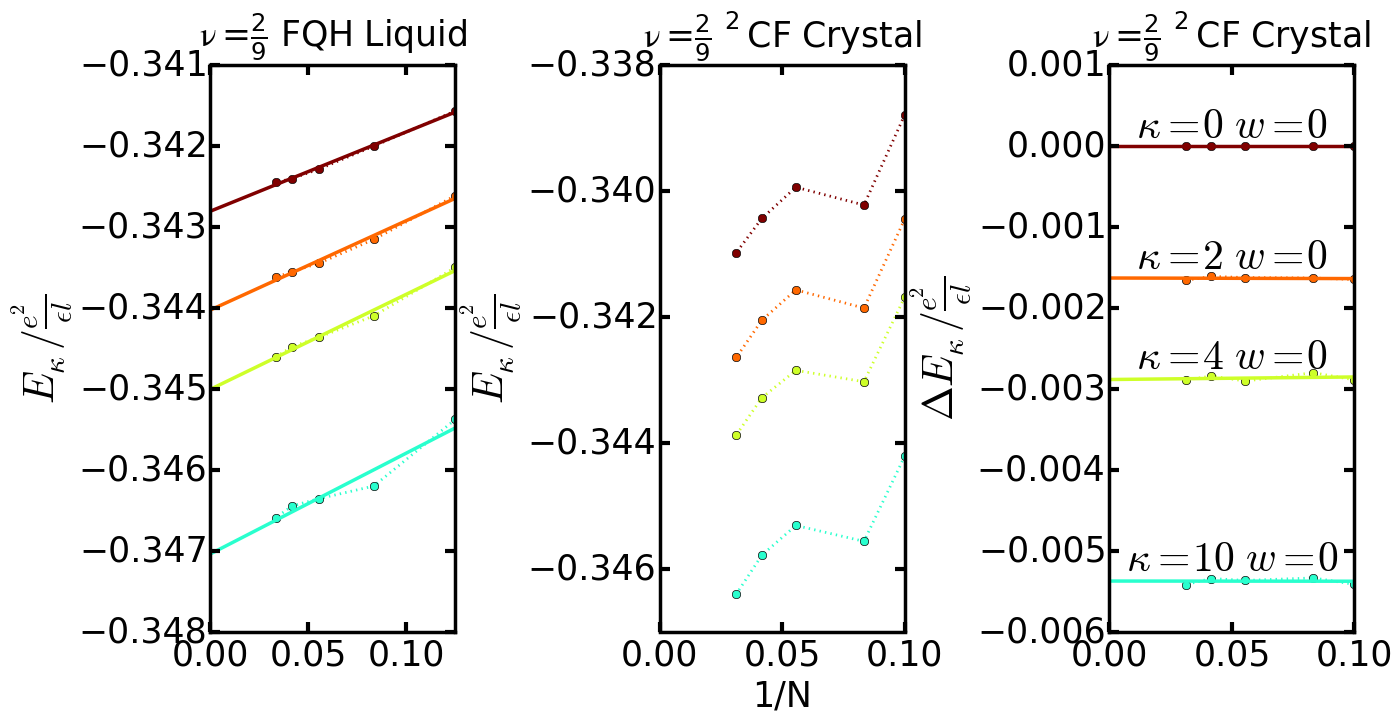}
\includegraphics[scale=0.3]{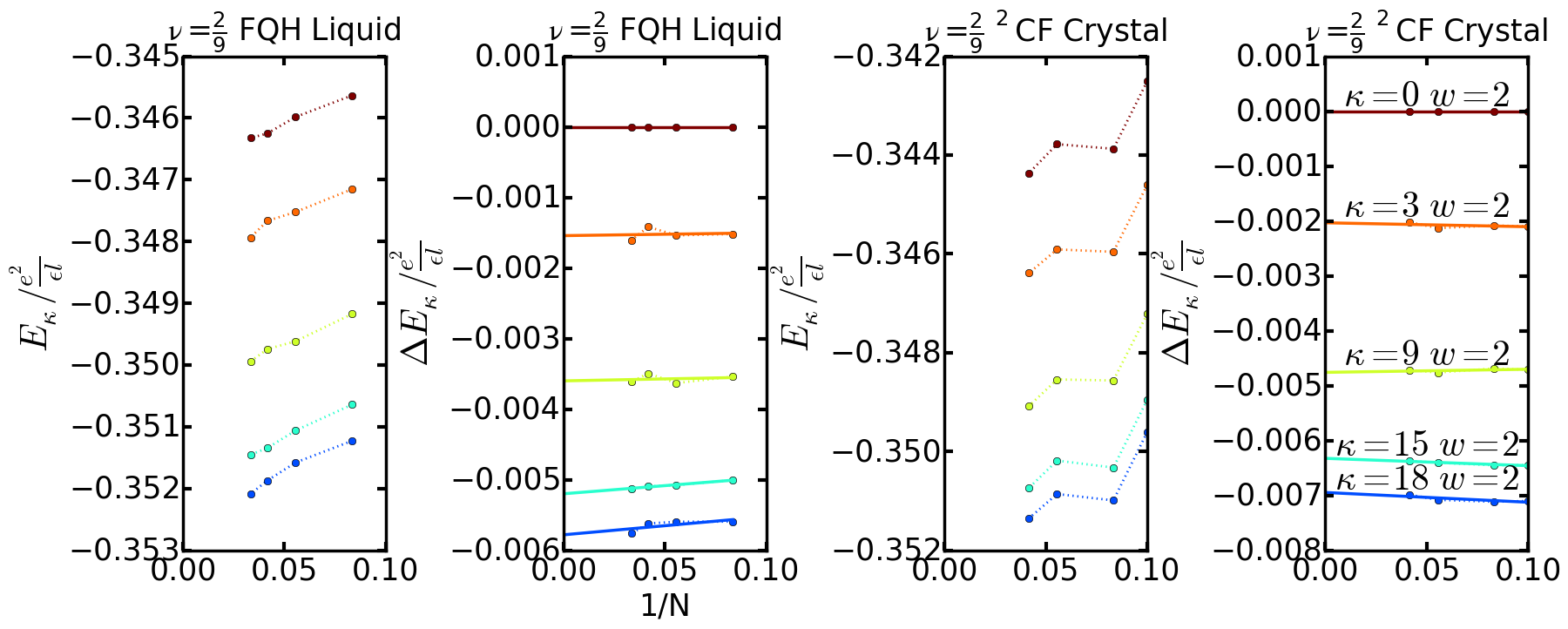}
\includegraphics[scale=0.3]{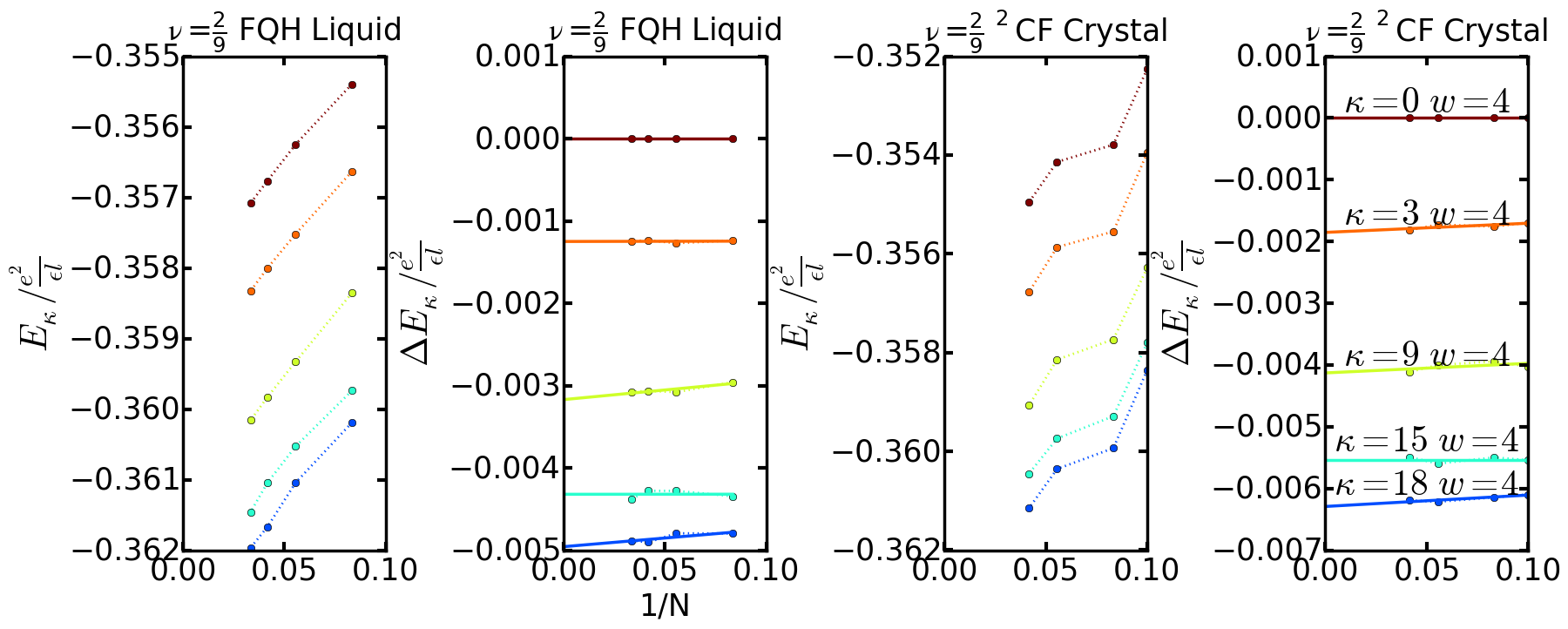}
\includegraphics[scale=0.3]{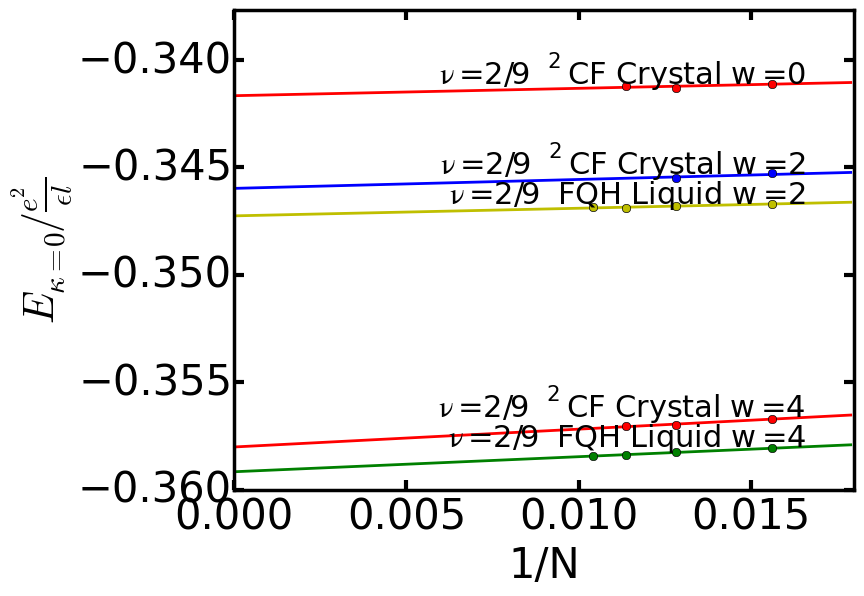}
\caption{
Same as in Fig.~\ref{1_3} but for $\nu=2/9$. Here the relevant crystal is the $^2$CF crystal.
}
\label{2_9}
\end{figure}

\begin{figure}
\includegraphics[scale=0.4]{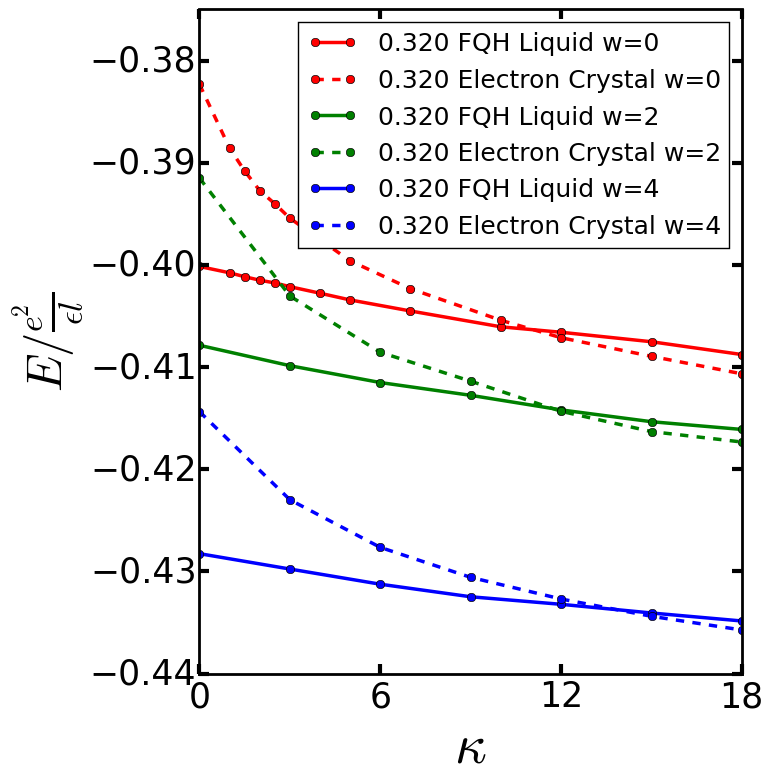}
\includegraphics[scale=0.4]{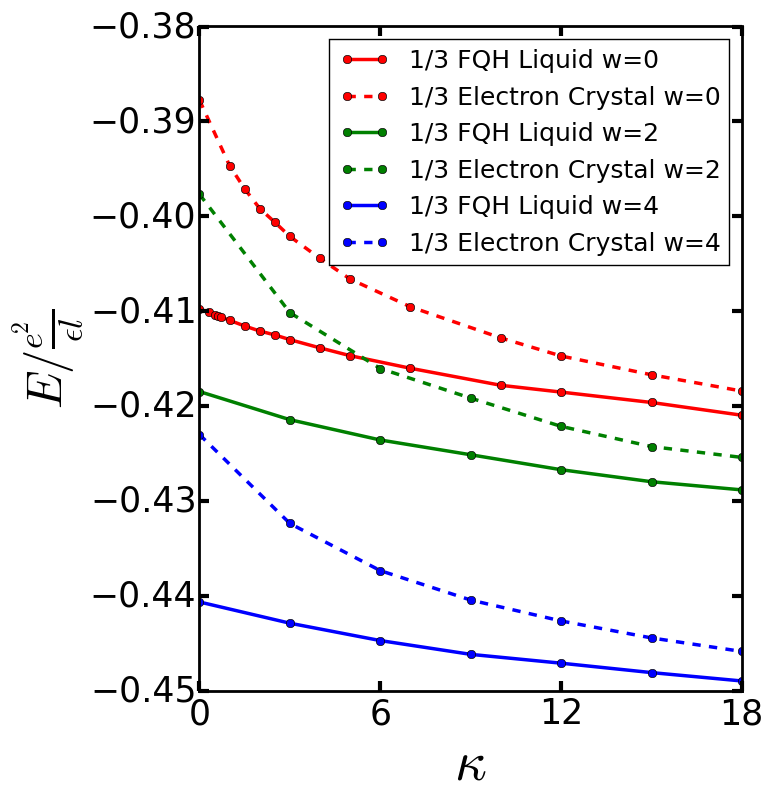}
\includegraphics[scale=0.4]{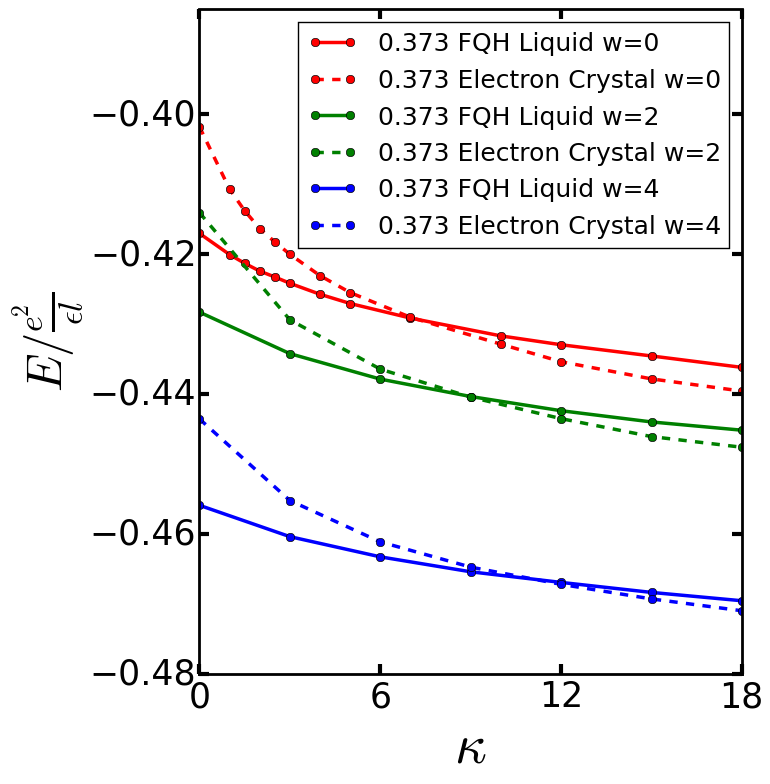}
\includegraphics[scale=0.4]{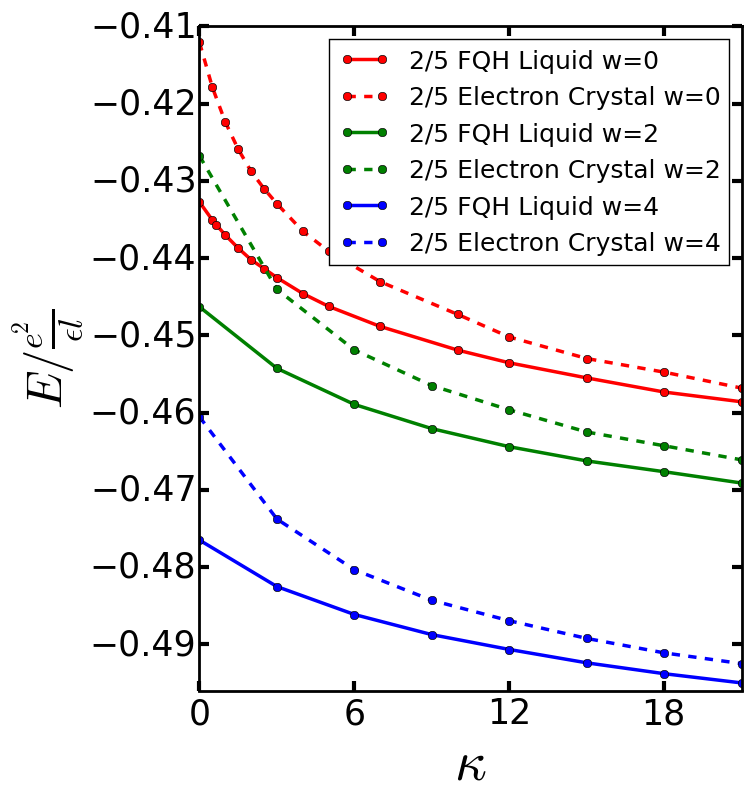}
\caption{
Thermodynamic energies as a function of $\kappa$ for filling factors $\nu=0.320$, $\nu=1/3$, $\nu=0.373$ and $\nu=2/5$, for quantum well widths $w=0$, $w=2l$ and $w=4l$.  At either $\nu={1\over3}$ or $\nu={2\over5}$, the FQH liquid remains the ground state for $\kappa$ up to $18$. At the nearby fillings $\nu=0.320$ and $\nu=0.373$, a level crossing transition takes place at $\kappa\approx 11$ and $\kappa\approx 8$, respectively, for $w=0$. Finite $w$ appears to push the transitions to slightly higher $\kappa$ values.
}
\label{1_3_kappa}
\end{figure}

\begin{figure}
\includegraphics[scale=0.4]{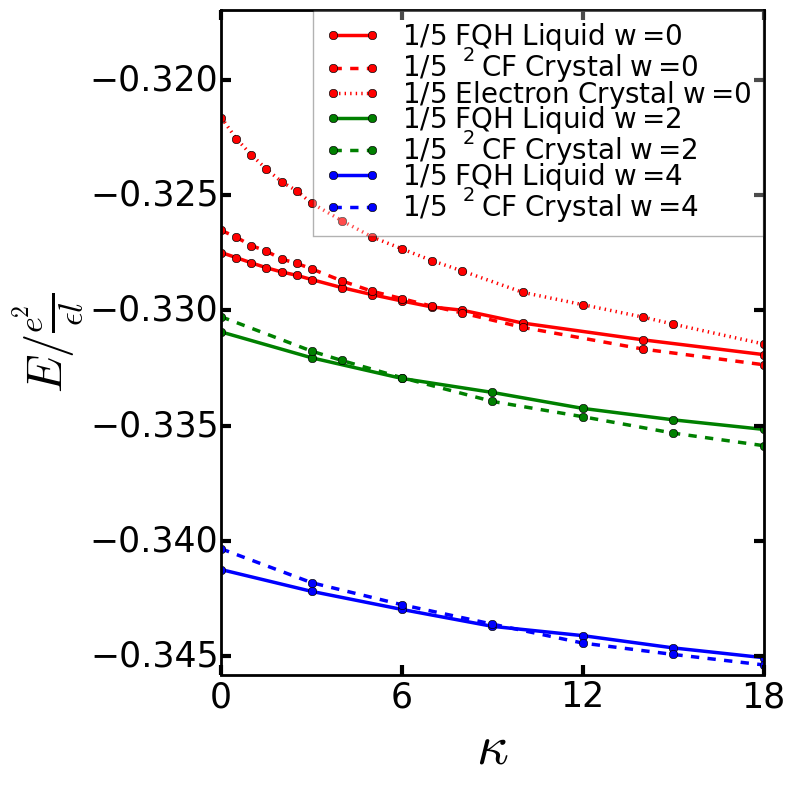}
\includegraphics[scale=0.4]{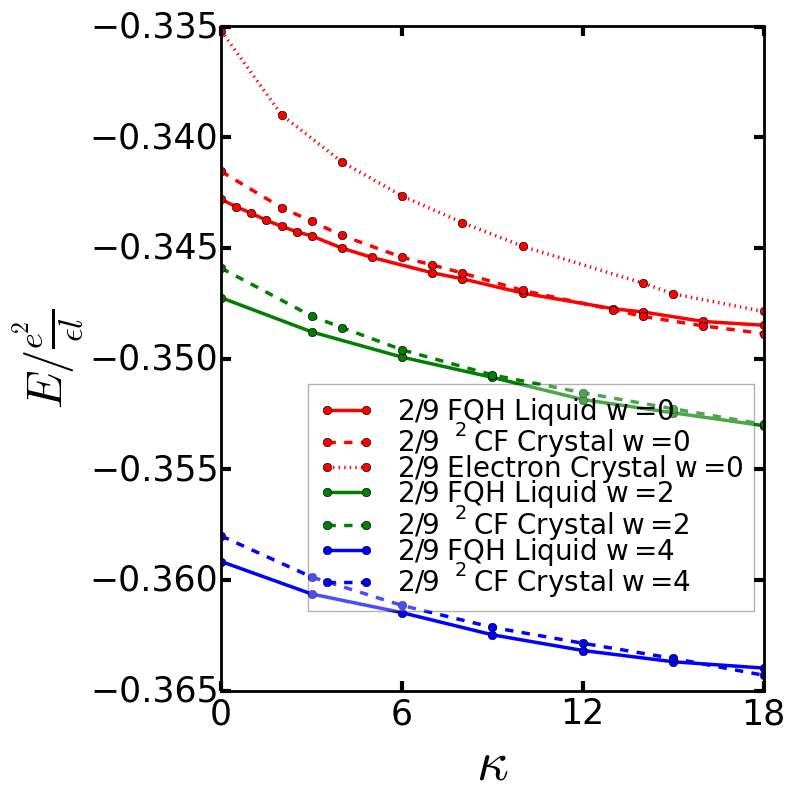}
\caption{
Energies as a function of $\kappa$ for $\nu={1\over5}$, ${2\over9}$. For finite widths, it is not possible to determine the level crossing point precisely, because the energies of the liquid and the crystal remain very close for a significant range of $\nu$. We also show energies of the {\em electron} crystal at $\nu={1\over5}$ and $\nu={2\over9}$ for $w=0l$, which always has a higher energy than the $^2$CF crystal at these two filling factors.
}
\label{1_5_kappa}
\end{figure}

\begin{figure}
\includegraphics[scale=0.5]{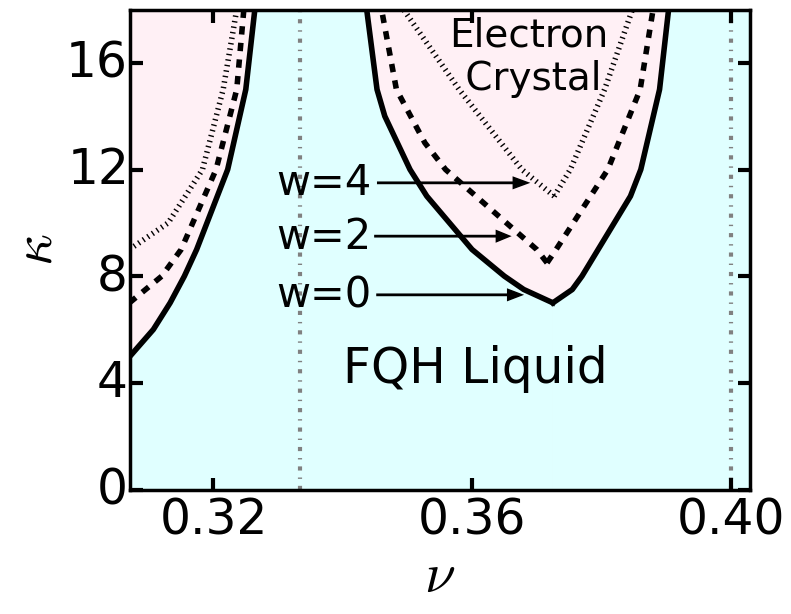}
\caption{
The phase diagram in a filling factor range including $\nu={1\over3}$ and $\nu={2\over5}$ for quantum well widths $w=0$, $w=2l$ and $w=4l$.
}
\label{phase_kappa}
\end{figure}

\end{widetext}
\end{document}